\documentclass[12pt, draftcls, onecolumn]{IEEEtran}
\usepackage{graphicx}
\usepackage{amsmath}
\usepackage{epsfig}
\usepackage{multirow}
\usepackage{amssymb}
\usepackage{algorithm}
\usepackage{algorithmic}
\usepackage{cases}
\usepackage{array}
\usepackage{booktabs}
\usepackage{multirow}

\ifCLASSINFOpdf
\else
\fi
\begin{document}
%
\title{Algebraic Soft Decoding of Reed-Solomon Codes Using Module Minimization}

\author{Jiongyue Xing, \IEEEmembership{Student Member, IEEE}, Li Chen, \IEEEmembership{Senior Member, IEEE}, \\Martin Bossert, \IEEEmembership{Fellow, IEEE}
\thanks{Jiongyue Xing and Li Chen are with the School of Electronics and Communication Engineering, Sun Yat-sen University, Guangzhou, P. R. China. Martin Bossert is with the Institute of Communications Engineering, Ulm University, Ulm, Germany. Email: xingjyue@mail2.sysu.edu.cn, chenli55@mail.sysu.edu.cn, martin.bossert@uni-ulm.de. This work is sponsored by the National Natural Science Foundation of China (NSFC) with project ID 61671486.}}


%


\maketitle

\begin{abstract}
The interpolation based algebraic decoding for Reed-Solomon (RS) codes can correct errors beyond half of the code's minimum Hamming distance. Using soft information, the algebraic soft decoding (ASD) further improves the decoding performance. This paper presents a unified study of two classical ASD algorithms in which the computationally expensive interpolation is solved by the module minimization (MM) technique. An explicit module basis construction for the two ASD algorithms will be introduced. Compared with Koetter's interpolation, the MM interpolation enables the algebraic Chase decoding and the Koetter-Vardy decoding perform less finite field arithmetic operations. Re-encoding transform is applied to further reduce the decoding complexity.
Computational cost of the two ASD algorithms as well as their re-encoding transformed variants are analyzed. This research shows re-encoding transform attributes to a lower decoding complexity by reducing the degree of module generators. Furthermore, Monte-Carlo simulation of the two ASD algorithms has been performed to show their decoding and complexity competency.
\end{abstract}


\IEEEpeerreviewmaketitle

\section{Introduction}
%
%
%
%
\IEEEPARstart{R}{eed}-Solomon (RS) codes are widely employed in data communications and storage systems, in which the well known Berlekamp-Massey (BM) decoding algorithm \cite{Berlekamp1968} \cite{Massey1969} is used. It is a syndrome based decoding that delivers at most one decoded message candidate. Hence, it is also called the unique decoding. Other RS unique decoding algorithms include the Euclidean algorithm \cite{SKHN1975} and the Welch-Berlekamp (WB) algorithm \cite{WB1986}. They all have an efficient running time but with a limited error-correction capability. For an $(n, k)$ RS code, where $n$ and $k$ are the length and dimension of the code, respectively, they can correct at most $\lfloor\frac{n - k}{2}\rfloor$ errors, i.e., half of the code's minimum Hamming distance. Assisted by soft information to perform the error-erasure decoding, the generalized minimum distance (GMD) decoding algorithm \cite{Kotter1996fast} and the modified WB algorithm \cite{Berlekamp1996} both achieve an improved decoding performance.

In late 90s, Sudan introduced an interpolation based algebraic decoding algorithm~\cite{Sudan1997} to correct errors beyond the above limit. But this improvement only applies to low rate codes. Guruswami and Sudan later improved it to decode all rate codes up to $n - \lfloor\sqrt{n(k - 1)}\rfloor -1$ errors~\cite{GS1999}. This is the so-called Guruswami-Sudan (GS) algorithm. Since this interpolation based decoding delivers a list of message candidates, it is also called the list decoding. It consists of two steps, interpolation that finds a minimum polynomial $Q(x,y)$ and factorization that finds $y$-roots of $Q(x,y)$. Interpolation dominates the decoding complexity. It is often implemented by Koetter's iterative polynomial construction algorithm~\cite{Koetter1996phd} \cite{NH2000}. It yields a Gr\"obner basis from which the minimum candidate is chosen as $Q(x,y)$. Algebraic soft decoding (ASD) was later introduced by Koetter and Vardy, namely the KV algorithm~\cite{KV2003}. It transforms soft received information into multiplicity information that defines the interpolation. It outperforms the GS algorithm. The other classical ASD algorithm is the algebraic Chase decoding (ACD) \cite{BK_LCC2010}. It constructs a number of decoding test-vectors whose formulation allows the following Koetter's interpolation to be performed in a low-complexity binary tree growth fashion. Also to decode beyond the half distance bound, power decoding was introduced by Schmidt \emph{et al.} \cite{SSB2010}. It achieves a similar error-correction capability as Sudan's algorithm for low rate codes by using multi-sequence shift-register synthesis. Another efficient list decoding algorithm was introduced by Wu \cite{Wu_list2008}. It utilizes the BM decoding output to construct $Q(x,y)$, leading to a lower interpolation multiplicity. Other complexity reducing approaches include the re-encoding transform \cite{KMV_reencoding2011}, the divide-and-conquer interpolation \cite{ma2004divide} and the progressive interpolation \cite{LSX_PASD2013}.

It has been reported that the interpolation problem can also be solved from the perspective of Gr\"{o}bner basis of module \cite{KF2002} \cite{Lee08}. It first formulates a basis of module which contains bivariate polynomials that interpolate all the prescribed points. Presenting it as a matrix over univariate polynomials, row operation further converts it into the Gr\"{o}bner basis defined under a weighted monomial order. Its minimum candidate will be $Q(x,y)$. We call this interpolation technique the module minimization (MM) which refers to the basis reduction process. Performing the GS and KV decoding using the MM interpolation has been investigated in \cite{Lee08} and \cite{Alekhnovich2005}. The basis reduction can be realized by either the Mulders-Storjohann (MS) algorithm \cite{Mulders03} or the more efficient Alekhnovich algorithm \cite{Alekhnovich2005}. Lee and O'Sullivan gave an explicit basis for the module and developed another basis reduction approach \cite{Lee08} \cite{Lee2006}. It is a variant of the MS algorithm that reduces the basis in a specific order. Nielsen and Zeh have also shown that the Alekhnovich algorithm is a divide-and-conquer variant of the MS algorithm \cite{JSR_Nielsen2014}. To achieve a better time complexity, Beelen and Brander \cite{BB2010key} employed the Alekhnovich algorithm on the basis proposed in \cite{Lee08}. Cohn and Heninger \cite{Cohn2015ideal} have demonstrated that a faster basis reduction approach could be realized by using the polynomial matrix multiplication \cite{GJV2003complexity}. Another fast basis reduction algorithm that utilizes structured linear algebra was proposed by Chowdhury \emph{et al.} \cite{CJNEV2015}. Recently, under the linearization framework, Jeannerod \emph{et al.} have proposed the fastest basis reduction approach using fast matrix computations \cite{jeannerod2017computing}. Further applying the re-encoding transform, Ma and Vardy have defined the explicit module generators for the KV algorithm \cite{Ma2007}. An MM based multi-trial GS decoding was introduced by Nielsen and Zeh \cite{JSR_Nielsen2014}. It is also a progressive RS decoding that gradually enlarges the decoding parameter until a satisfied decoding outcome is produced. Performing ACD using MM has been presented by Chen and Bossert \cite{Li_ACDMM2016}. The MM interpolation approach has also been generalized to perform Wu's list decoding and power decoding in \cite{BHNW2013} and \cite{Rosenkilde2018power}, respectively. Meanwhile, the interpolation problem can be reformulated into a system of key equations with univariate polynomials \cite{Zeh2011interpolation} and it can also be solved by the MM technique \cite{Nielsen2013phd}.

However, despite its complexity advantage over Koetter's interpolation, performing the ASD algorithms using MM interpolation has so far received light attention in the community. Many practical aspects of this approach demand a more comprehensive understanding. For example, engineers prefer to conceive a more systematic module formulation especially when the re-encoding transform is applied. The exact complexity reduction yielded by the MM technique and the re-encoding transform is still unknown for practical codes. It is also beneficial to know the performance and complexity competency between the ACD and the KV algorithms.
Therefore, this paper presents a more comprehensive and unified view of the above two algorithms, both of which utilize the MM interpolation. From this point onward, they are named the ACD-MM algorithm and the KV-MM algorithm, respectively. Our contributions can be outlined as follows.

$\bullet$~We will give a unified description on how to use the MM technique to solve the interpolation problem in the two classical ASD algorithms. This is especially challenging when formulating a module basis for the KV-MM algorithm. We will show the formulation is underpinned by the point enumeration process.

$\bullet$~We will further present the re-encoding transformed variants of the ACD-MM and the KV-MM algorithms. Our research reveals this transform helps reduce the degree of module generators, which leads to a simpler basis reduction process.

$\bullet$~Complexity of the ACD-MM and the KV-MM algorithms will be analyzed. We will show the MM interpolation and its re-encoding transform are more effective in yielding a low complexity for high rate codes, which is welcomed by practice.

$\bullet$~Finally, we will provide simulation results on complexity and decoding performance of the two algorithms. Our results show the MM interpolation requires less finite field arithmetic operations than Koetter's interpolation. The re-encoding transform can further reduce the MM complexity. Performance and complexity of the two ASD algorithms will be compared, providing more practical insights.

The rest of this paper is organized as follows. Section II introduces RS codes and the GS decoding using MM. Section III introduces the ACD-MM algorithm and its re-encoding transformed variant. Section IV introduces the KV-MM algorithm and its re-encoding transformed variant. Section V analyzes complexity of the two ASD algorithms and Section VI presents our simulation results. Finally, Section VII concludes the paper.

\section{RS Codes and GS Decoding Using MM}
This section introduces the prerequisites of the paper, including the RS encoding and the GS decoding based on the MM interpolation.

\subsection{RS Codes}
Let $\mathbb{F}_{q} = \{\sigma_0,\sigma_1,\ldots, \sigma_{q-1}\}$ denote the finite field of size $q$, and $\mathbb{F}_{q}[x]$ and $\mathbb{F}_{q}[x, y]$ denote the univariate and the bivariate polynomial rings defined over $\mathbb{F}_{q}$, respectively. For an $(n,k)$ RS code, where $n = q - 1$, the message polynomial $f(x)\in\mathbb{F}_{q}[x]$ is
\begin{equation}\label{message}
f(x) = f_{0} + f_{1}x + \cdot\cdot\cdot + f_{k - 1}x^{k - 1},
\end{equation}
where $f_{0}, f_{1}, \ldots, f_{k - 1}$ are message symbols. The codeword $\underline{c} = (c_{0}, c_{1}, \ldots, c_{n - 1})\in\mathbb{F}_{q}^{n}$ can be generated by
\begin{equation}\label{codeword}
\underline{c} = (f(\alpha_{0}), f(\alpha_{1}), \ldots, f(\alpha_{n - 1})),
\end{equation}
where $\alpha_{0}, \alpha_{1}, \ldots, \alpha_{n - 1}$ are $n$ distinct nonzero elements of $\mathbb{F}_{q}$. They are called the code locators.

\subsection{GS Decoding Using MM}
Let $\underline{\omega} = (\omega_{0}, \omega_{1}, \ldots, \omega_{n - 1})\in\mathbb{F}_{q}^{n}$ denote the received word. The GS decoding algorithm~\cite{GS1999} consists of two steps, interpolation and root-finding. First, the $n$ interpolation points $(\alpha_{0}, \omega_{0})$, $(\alpha_{1}, \omega_{1})$, $\ldots$ , $(\alpha_{n - 1}, \omega_{n - 1})$ are formed.
The Hamming distance between $\underline{c}$ and $\underline{\omega}$ is $d_{\text{H}}(\underline{c}, \underline{\omega}) = | \{j \mid c_{j} \neq \omega_{j}, \forall j\}|$.
Given a polynomial $Q(x,y)=\sum_{a,b}Q_{ab}x^ay^b\in\mathbb{F}_q[x,y]$, its monomials $x^ay^b$ can be organized under the $(\mu,\nu)$-revlex order
\footnote{The $(\mu,\nu)$-weighted degree of $x^ay^b$ is $\deg_{\mu,\nu}x^ay^b=\mu a+\nu b$. Given $x^{a_1}y^{b_1}$ and $x^{a_2}y^{b_2}$, it is claimed $\text{ord}(x^{a_1}y^{b_1})<\text{ord}(x^{a_2}y^{b_2})$, if $\deg_{\mu,\nu}x^{a_1}y^{b_1}<\deg_{\mu,\nu}x^{a_2}y^{b_2}$, or $\deg_{\mu,\nu}x^{a_1}y^{b_1}=\deg_{\mu,\nu}x^{a_2}y^{b_2}$ and $b_1<b_2$.}.
Let $x^{a'}y^{b'}$ denote the leading monomial of $Q$ where $Q_{a'b'}\neq0$, the $(\mu,\nu)$-weighted degree of $Q$ is $\deg_{\mu,\nu}Q=\deg_{\mu,\nu}x^{a'}y^{b'}$. Furthermore, given polynomials $Q_1$ and $Q_2$ with leading monomials $x^{a_1}y^{b_1}$ and $x^{a_2}y^{b_2}$, respectively, $Q_1<Q_2$ if $\text{ord}(x^{a_1}y^{b_1})<\text{ord}(x^{a_2}y^{b_2})$.
The following GS decoding theorem can be introduced.

\textit{\textbf{Theorem 1}} \cite{GS1999}. For an $(n,k)$ RS code, let $Q$ $\in$ $\mathbb{F}_{q}[x, y]$ denote a polynomial that interpolates the $n$ points with a multiplicity of $m$. If $m(n - d_{\text{H}}$($\underline{c}$, $\underline{\omega}))$ $>$ $\deg_{1, k - 1}Q(x, y)$, $Q(x, f(x)) = 0$.

Therefore, interpolation finds $Q$ with the minimum $(1, k - 1)$-weighted degree, where the message $f(x)$ can be recovered by finding the $y$-roots of $Q$ \cite{RR_factorization2000}. Hence, the maximum decoding output list size (OLS) is $\deg_y Q$. Let $l=\deg_y Q$ and it is the decoding parameter in this paper. Note that in the GS algorithm, $m \leq l$ \cite{GS1999} \cite{Nielsen2013phd}.

\textit{\textbf{Definition I.}} Let $\underline{\xi} = (\xi_{0}(x), \xi_{1}(x), \ldots, \xi_{l}(x))$ denote a
vector over $\mathbb{F}_{q}[x]$, the degree of $\underline{\xi}$ is
\begin{equation}\label{rowdegree}
\deg\underline{\xi} = \max\{\deg\xi_{\tau}(x),\forall \tau\}.
\end{equation}
The leading position (LP) of $\underline{\xi}$ is
\begin{equation}\label{lp}
\text{LP}(\underline{\xi}) = \max\{\tau \mid \deg\xi_{\tau}(x) =  \deg\underline{\xi}\}.
\end{equation}
Since $\xi_{\tau}(x) = \xi_{\tau}^{(0)} + \xi_{\tau}^{(1)}x + \cdots + \xi_{\tau}^{(\deg\xi_{\tau}(x))}x^{\deg\xi_{\tau}(x)}$, the leading term (LT) of $\xi_{\tau}(x)$ is
\begin{equation}\label{lt}
\text{LT}(\xi_{\tau}(x)) = \xi_{\tau}^{(\deg\xi_{\tau}(x))}x^{\deg\xi_{\tau}(x)}.
\end{equation}

\textit{\textbf{Definition II.}} Given a matrix $\mathcal{V}$ over $\mathbb{F}_q[x]$, we denote its row-$t$ by $\mathcal{V}|_t$ and its entry of row-$t$ column-$\tau$ by $\mathcal{V}|_t^{(\tau)}$. Furthermore, the degree of $\mathcal{V}$ is
\begin{equation}\label{matrixdegree}
\deg \mathcal{V} = \sum_t \deg \mathcal{V}|_t.
\end{equation}

\textit{\textbf{Definition III.}} Let $\mathcal{M}_{l}$ denote the space of all bivariate polynomials over $\mathbb{F}_{q}[x, y]$ that interpolate all prescribed points with their multiplicity and have a maximum $y$-degree of $l$.

$\mathcal{M}_{l}$ can be viewed as an $\mathbb{F}_q[x]$-module \cite{Lee08}. In GS algorithm with a multiplicity of $m$ and a decoding OLS of $l$, the explicit basis of $\mathcal{M}_{l}$ can be constructed by
\begin{equation}\label{G_poly}
G(x) = \prod\limits_{j = 0}^{n - 1}(x - \alpha_{j})
\end{equation}
and
\begin{equation}\label{R_poly}
R(x) = \sum\limits_{j = 0}^{n - 1}\omega_{j}\Phi_{j}(x),
\end{equation}
where
\begin{equation}~\label{lagrange}
\Phi_{j}(x) = \prod_{j' = 0, j' \neq j}^{n - 1} \frac{x - \alpha_{j'}}{\alpha_{j} - \alpha_{j'}}
\end{equation}
is the Lagrange basis polynomial. It satisfies $\Phi_j(\alpha_j) = 1$ and $\Phi_j(\alpha_{j'}) = 0,\forall j'\neq j$. As a result, $R(\alpha_j) = \omega_j, \forall j$. $\mathcal{M}_{l}$ can be generated as an $\mathbb{F}_q[x]$-module by the following $l + 1$ polynomials \cite{Lee08} \cite{Nielsen2013phd}
\begin{equation}\label{Pt_1}
P_{t}(x, y) = G(x)^{m - t}(y - R(x))^{t}, \text{if } 0 \leq t \leq m,
\end{equation}
\begin{equation}\label{Pt_2}
P_{t}(x, y) = y^{t - m}(y - R(x))^{m}, \text{if } m < t \leq l.
\end{equation}
Note that $P_t(\alpha_j, \omega_j) = 0,\forall(t, j)$ and the total degree of $y - R(x)$ and $G(x)$ is $m$. Since any element of $\mathcal{M}_l$ can be presented as an $\mathbb{F}_q[x]$-linear combination of $P_t(x,y)$, (\ref{Pt_1}) and (\ref{Pt_2}) form a basis $\mathcal{B}_l$ of module $\mathcal{M}_l$ \cite{JSR_Nielsen2014}. Moreover, since $P_{t}(x, y) = \sum_{\tau \leq t} P_{t}^{(\tau)}(x)y^{\tau}$ where $P_{t}^{(\tau)}(x) \in \mathbb{F}_{q}[x]$, $\mathcal{B}_l$ can be presented as an $(l + 1) \times (l + 1)$ matrix over $\mathbb{F}_q[x]$ by letting $\mathcal{B}_{l}|_t^{(\tau)} = P_t^{(\tau)}(x)$. Each row of the matrix corresponds to a bivariate polynomial of $\mathcal{B}_l$. Afterwards, the basis reduction, e.g., the MS algorithm \cite{Mulders03}, transforms $\mathcal{B}_l$  into the Gr\"obner basis of $\mathcal{M}_l$. The following Proposition defines the Gr\"obner basis of $\mathcal{M}_l$.


\textit{\textbf{Proposition 1}} \cite{Lee2006}. Assume that $\{g_t\in\mathbb{F}_q[x,y],0\leq t\leq l\}$ generates module $\mathcal{M}_l$. Under the $(\mu,\nu)$-revlex order, if the $y$-degree of the leading nominal of each polynomial $g_t$ is different, $\{g_t\in\mathbb{F}_q[x,y],0\leq t\leq l\}$ is a Gr\"obner basis of $\mathcal{M}_l$.

%
%
%

\textit{\textbf{Definition IV}}~\cite{Mulders03}. Given a square matrix $\mathcal{V}$ over $\mathbb{F}_q[x]$, if any of its two rows $\mathcal{V}|_t$ and $\mathcal{V}|_{t'}$ exhibit $\text{LP}(\mathcal{V}|_t)\neq\text{LP}(\mathcal{V}|_{t'})$, it is in the \textit{weak~Popov~form}.

Let $\mathcal{D}_{\beta, l} = \text{diag}(1, x^{\beta}, \ldots, x^{l\beta})$ denote the diagonal matrix of size $(l + 1) \times (l + 1)$ and $\beta$ is an integer. In decoding an $(n,k)$ RS code, performing the mapping of
\begin{equation}\label{A_ml_1_map}
\mathcal{A}_{l} = \mathcal{B}_{l} \cdot \mathcal{D}_{k - 1,l}
\end{equation}
enables $\deg\mathcal{A}_{l}|_t = \deg_{1, k - 1}P_{t}(x, y)$. The MS algorithm further reduces $\mathcal{A}_{l}$ into the weak Popov form $\mathcal{A}_{l}^{\prime}$ as follows. Find two rows of $\mathcal{A}_l$ such that $\deg\mathcal{A}_{l}|_{t} \leq \deg\mathcal{A}_{l}|_{t'}$ and $\text{LP}(\mathcal{A}_{l}|_{t}) = \text{LP}(\mathcal{A}_{l}|_{t'})$, and perform
\begin{equation}\label{MS algorithm}
\mathcal{A}_{l}|_{t'}=\mathcal{A}_{l}|_{t'} - \frac{\text{LT}(\mathcal{A}_{l}|_{t'}^{\text{LP}(\mathcal{A}_{l}|_{t'})})}{\text{LT}(\mathcal{A}_{l}|_t^{\text{LP}(\mathcal{A}_{l}|_{t})})}
         \cdot\mathcal{A}_{l}|_{t}.
\end{equation}
Repeat this operation until the weak Popov form $\mathcal{A}_l'$ is reached. Performing demapping of
\begin{equation}\label{D_ml_1_demap}
\mathcal{B}_{l}^{\prime} = \mathcal{A}_{l}^{\prime} \cdot \mathcal{D}_{-(k - 1),l},
\end{equation}
$\mathcal{B}_l'$ is a Gr\"obner basis of $\mathcal{M}_{l}$. Based on $\mathcal{B}_l'|_t$, we can construct $P_t'(x,y)$ by letting $P_t'^{(\tau)}(x)=\mathcal{B}_l'|_t^{(\tau)}$. Since $\deg_{1,k-1}P_t'(x,y)=\deg\mathcal{A}_l'|_t=\deg\mathcal{A}_l'|_t^{(\text{LP}(\mathcal{A}_l'|_t))}=\deg P_t'^{(\text{LP}(\mathcal{A}_l'|_t))}(x)+(k-1)\cdot\text{LP}(\mathcal{A}_l'|_t)$, when $\mathcal{A}_l'$ is in the weak Popov form, $y$-degree of each polynomial's leading monomial, i.e., $\text{LP}(\mathcal{A}_l'|_t)$, is different. Based on Proposition 1, $\mathcal{B}_l'$ is a Gr\"obner basis. Let $\mathcal{A}_{l}^{\prime}|_{t^*}$ denote the row that has the minimum degree, the interpolated polynomial $Q(x, y)$ can be constructed from $\mathcal{B}_{l}'|_{t^*}$ by letting
\begin{equation}\label{QfromM}
Q^{(\tau)}(x) = \mathcal{B}_{l}'|_{t^*}^{(\tau)},\forall\tau.
\end{equation}
Finally, determine the $y$-roots of $Q$ using the recursive coefficient search algorithm \cite{RR_factorization2000}.


\section{The ACD-MM Algorithm}
This section introduces the ACD-MM algorithm. It constructs a number of decoding test-vectors based on the reliability matrix. The GS decoding will be further performed on each test-vector. We will also introduce its re-encoding transformed variant.

\subsection{From Reliability Matrix to Test-Vectors}
Assume codeword $\underline{c} = (c_0, c_1, \ldots, c_{n-1})$ is transmitted through a memoryless channel and $\underline{\mathsf{r}} = (\mathsf{r}_0, \mathsf{r}_1, \ldots, \mathsf{r}_{n - 1}) \in \mathbb{R}^n$ is the received symbol vector, where $\mathbb{R}$ denotes the channel output alphabet. The channel observation is represented by a reliability matrix $\mathbf{\Pi}$ whose entries are the \emph{a posteriori} probability (APP)~\footnote{Assume that $\Pr[c_j = \sigma_i]=\frac{1}{q},\forall i$.} defined as
\begin{equation}\label{pi_ij}
\pi_{ij} = \Pr[c_j = \sigma_i~|~\mathsf{r}_j],~\text{for}~0 \leq i \leq q-1~\text{and}~0 \leq j \leq n - 1.
\end{equation}
With matrix $\mathbf{\Pi}$, let
\begin{equation} \label{1pi2pi}
i_{j}^{\text{I}}=\arg\max\{\pi_{ij},\forall i\}~\text{and}~i_{j}^{\text{II}}=\arg\max\{\pi_{ij},\forall i~\text{and}~i\neq i_{j}^{\text{I}}\},
\end{equation}
respectively, where function $\arg\max$ returns index $i$. The two most likely decisions for $c_j$ are
\begin{equation}\label{r_12}
r_{j}^{\text{I}}=\sigma_{i_{j}^{\text{I}}}~\text{and}~r_{j}^{\text{II}}=\sigma_{i_{j}^{\text{II}}}.
\end{equation}
Define the symbol wise reliability metric as~\cite{BK_LCC2010}
\begin{equation}\label{gamma}
\gamma_{j} = \frac{\pi_{i_{j}^{\text{II}}j}}{\pi_{i_{j}^{\text{I}}j}},
\end{equation}
where $\gamma_{j} \in (0, 1)$. With $\gamma_{j} \rightarrow 0$, the decision on $c_{j}$ is more
reliable, and vise versa. By sorting the $n$ reliability metrics in an ascending order, we obtain a refreshed symbol index sequence $j_{0}, j_{1},\ldots, j_{n - 1}$. It indicates $\gamma_{j_{0}}<\gamma_{j_{1}}<\cdots<\gamma_{j_{n - 1}}$. Choose $\eta~(\eta < n)$ least reliable symbols that can be realized as either $r_j^{\text{I}}$ or $r_j^{\text{II}}$. For the remaining $n - \eta$ reliable symbols, they will be realized as $r_j^{\text{I}}$. We can formulate $2^{\eta}$ interpolation test-vectors which can be generally written as
\begin{equation}\label{tv}
\underline{r}_{u} = (r_{j_{0}}^{(u)}, r_{j_{1}}^{(u)}, \ldots, r_{j_{k - 1}}^{(u)}, r_{j_{k}}^{(u)}, \ldots, r_{j_{n - 1}}^{(u)}),
\end{equation}
where $u = 1, 2, \ldots, 2^{\eta}$, $r_j^{(u)} = r_{j}^{\text{I}}$ for $j = j_{0}, j_{1}, \ldots, j_{n - \eta - 1}$, and $r_j^{(u)} = r_{j}^{\text{I}}~\text{or}~r_{j}^{\text{II}}$ for $j = j_{n - \eta}, j_{n - \eta + 1}, \ldots, j_{n - 1}$.

Other less likely decisions can also be considered for the unreliable symbols. As a result, the number of test-vectors increases exponentially. However, our research has shown little performance gain can be achieved by considering more than two decisions.

\subsection{Module Formulation and Minimization}

For each test-vector, the MM based GS decoding that is described in Section II.B will be performed. In particular, given a test-vector $\underline{r}_u$, polynomial $R(x)$ of (\ref{R_poly}) becomes
\begin{equation} \label{Ru_poly}
R_u(x) = \sum_{j = 0}^{n - 1}r_j^{(u)} \Phi_j(x).
\end{equation}
Consequently, $R_u(\alpha_j) = r_j^{(u)},\forall j$. Module $\mathcal{M}_l$ can be formed using the $l+1$ generators of (\ref{Pt_1}) (\ref{Pt_2}), in which $R(x)$ is replaced by $R_u(x)$.

Note that the MM interpolation for each test-vector is independent. They can be performed in parallel, leveraging the decoding latency to that of a single GS decoding event. This is an advantage over the low-complexity Chase (LCC) algorithm that employs Koetter's interpolation in a binary tree growth fashion \cite{BK_LCC2010}.

The ACD-MM algorithm is summarized as follows. Note that $\hat{f}(x)$ denotes the decoding estimation of $f(x)$.

\begin{algorithm}[htbp]
\caption{The ACD-MM Algorithm}
\label{alg:The ACD-MM Algorithm}
 \textbf{Input:} $\mathbf{\Pi},\eta,m,l$;\\
 \textbf{Output:} $\hat{f}(x)$;
 \begin{itemize}
 \item[\textbf{ 1:}]Determine metrics $\gamma_j$ as in (\ref{gamma});
 \item[\textbf{ 2:}]Formulate $2^{\eta}$ test-vectors $\underline{r}_u$ as in (\ref{tv});
 \item[\textbf{ 3:}]\textbf{For} each test-vector $\underline{r}_u$ \textbf{do}
 \item[\textbf{ 4:}]\quad Formulate $\mathcal{B}_l$ by (\ref{Pt_1}) (\ref{Pt_2}) and map to $\mathcal{A}_l$ by (\ref{A_ml_1_map});
 \item[\textbf{ 5:}]\quad Reduce $\mathcal{A}_l$ into $\mathcal{A}_l'$ and demap it to $\mathcal{B}_l'$ by (\ref{D_ml_1_demap});
 \item[\textbf{ 6:}]\quad Construct $Q$ as in (\ref{QfromM});
 \item[\textbf{ 7:}]\quad Factorize $Q$ to find $\hat{f}(x)$;
 \item[\textbf{ 8:}]\textbf{End for}
\end{itemize}
\end{algorithm}

\subsection{The Re-encoding Transformed ACD-MM}
Re-encoding transforms the test-vectors so that they have at least $k$ zero symbols. This will reduce the $x$-degree of module generators, resulting in a simpler basis reduction process. Let $\Theta = \{j_0, j_1, \ldots, j_{k-1}\}$ denote the index set of the $k$ most reliable symbols. Its complementary set $\bar{\Theta}=\{j_{k},j_{k+1},\ldots,j_{n-1}\}$. We restrict $\eta \leq n - k$ so that the test-vectors would share at least $k$ common symbols $r_{j_0}^{\text{I}}, r_{j_1}^{\text{I}}, \ldots, r_{j_{k-1}}^{\text{I}}$. The $k$ re-encoding points are $(\alpha_j,r_j^{\text{I}})$ where $j\in\Theta$. The re-encoding polynomial is
\begin{equation}\label{H_poly}
H_{\Theta}(x) = \sum\limits_{j \in \Theta}r_{j}^{\text{I}}\prod_{j' \in \Theta, j' \neq j}\frac{x - \alpha_{j'}}{\alpha_{j} - \alpha_{j'}},
\end{equation}
where $H_{\Theta}(\alpha_j) = r_j^{\text{I}},\forall j \in \Theta$. All test-vectors $\underline{r}_{u}$ are transformed by
\begin{equation}\label{tv_transform}
\underline{r}_{u} \mapsto \underline{z}_{u} : z_{j}^{(u)} = r_{j}^{(u)} - H_{\Theta}(\alpha_{j}), \forall j.
\end{equation}
Consequently, the transformed test-vectors can be generally written as
\begin{equation}\label{transform_tv}
\underline{z}_{u} = (0, 0, \ldots, 0, z_{j_{k}}^{(u)}, \ldots, z_{j_{n - 1}}^{(u)}).
\end{equation}

With a transformed test-vector $\underline{z}_{u}$, polynomial $R_u(x)$ of (\ref{Ru_poly}) is redefined as
\begin{equation}\label{R_poly_mod}
R_u(x) = \sum\limits_{j = 0}^{n - 1}z_{j}^{(u)}\Phi_{j}(x).
\end{equation}
Since $z_{j}^{(u)} = 0, \forall j \in \Theta$,
\begin{equation}\label{V_poly}
V(x) = \prod\limits_{j \in \Theta}(x - \alpha_{j})
\end{equation}
becomes the GCD for both $G(x)$ and the above $R_u(x)$. Therefore, given $\underline{z}_{u}$, we define
\begin{equation}\label{G_poly_1}
\tilde{G}(x) = \frac{G(x)}{V(x)} = \prod\limits_{j \in \bar{\Theta}}(x - \alpha_{j})
\end{equation}
and
\begin{equation}\label{R_poly_1}
\tilde{R}_u(x) = \frac{R_u(x)}{V(x)} = \sum\limits_{j \in \bar{\Theta}}\frac{z_{j}^{(u)}}{\varpi_{j}}~\prod\limits_{j' \in \bar{\Theta}, j' \neq j}(x - \alpha_{j'}),
\end{equation}
where $\varpi_{j} = \prod_{j' = 0, j' \neq j}^{n - 1}(\alpha_{j} - \alpha_{j'})$.

The following Lemma further reveals the property of the module generators.

\textit{\textbf{Lemma 2}} \cite{Li_ACDMM2016}. Given a transformed test-vector $\underline{z}_{u}$ and a multiplicity $m$, $V(x)^{m}|P_{t}(x, yV(x))$.


Therefore, we can define the following bijective mapping
\begin{equation}\label{Ptu_map}
\begin{aligned}
\varphi:~~~~ \mathcal{M}_l &\rightarrow \mathbb{F}_q[x,y]\\
P_t(x,y) &\mapsto V(x)^{-m}P_{t}(x, yV(x)),
\end{aligned}
\end{equation}
where $\varphi$ is an isomorphism between $\mathcal{M}_l$ and $\varphi(\mathcal{M}_l)$. Polynomials of $\varphi(\mathcal{M}_l)$ have lower $x$-degree than those of $\mathcal{M}_l$. Since the MS basis reduction algorithm performs linear combination between its polynomials, the mapping of (\ref{Ptu_map}) will result in a simpler basis reduction process. Based on (\ref{Pt_1}) and (\ref{Pt_2}), we can further obtain the generators of $\varphi(\mathcal{M}_l)$ as
\begin{equation}\label{Ptu_1_map}
\tilde{P}_{t}(x, y) = \tilde{G}(x)^{m - t}(y - \tilde{R}_u(x))^{t},~\text{if}~0 \leq t \leq m,
\end{equation}
\begin{equation}\label{Ptu_2_map}
\tilde{P}_{t}(x, y) = (yV(x))^{t - m}(y - \tilde{R}_u(x))^{m},~\text{if}~m < t \leq l.
\end{equation}
They form a basis $\tilde{\mathcal{B}}_l$ of $\varphi(\mathcal{M}_l)$. Again, $\tilde{\mathcal{B}}_l$ can be presented as a square matrix over $\mathbb{F}_q[x]$.
Note that polynomials are now arranged under the $(1, -1)$-revlex order \cite{KMV_reencoding2011}. However, performing $\mathcal{A}_{l} = \tilde{\mathcal{B}}_l\cdot \mathcal{D}_{-1, l}$ will cause some of the basis entries leaving $\mathbb{F}_{q}[x]$. Alternatively, we let $\tilde{\mathcal{D}}_{\beta, l}=\text{diag}(x^{l\beta},x^{(l-1)\beta},\ldots,1)$ and $\mathcal{A}_l$ will be generated by
\begin{equation}\label{mapping_re_encoding}
\mathcal{A}_{l} = \tilde{\mathcal{B}}_l\cdot \tilde{\mathcal{D}}_{1, l},
\end{equation}
so that $\deg\mathcal{A}_{l}|_t = \deg_{1, -1}\tilde{P}_{t}(x, y) + l$. The MS algorithm will then reduce $\mathcal{A}_{l}$ into the weak Popov form $\mathcal{A}_l'$. Demap it by
\begin{equation}\label{demapping_re_encoding}
\tilde{\mathcal{B}}_l'=\mathcal{A}_l'\cdot\tilde{\mathcal{D}}_{-1, l}
\end{equation}
and polynomial $\tilde{Q}(x,y)$ can be retrieved from $\tilde{\mathcal{B}}_l'$ as in (\ref{QfromM}). Finally, the interpolated polynomial $Q$ can be constructed by
\begin{equation}\label{Q_restore}
Q(x, y) = V(x)^{m}\tilde{Q}\Big(x,\frac{y}{V(x)}\Big).
\end{equation}
If $f'(x)$ is a $y$-root of $Q$, $\hat{f}(x)$ is estimated by $\hat{f}(x)=f'(x)+H_{\Theta}(x)$.

The re-encoding transformed ACD-MM algorithm is summarized as in Algorithm \ref{alg:The Re-encoding Transformed ACD-MM Algorithm}.

\begin{algorithm}[htbp]
\caption{The Re-encoding Transformed ACD-MM Algorithm}
\label{alg:The Re-encoding Transformed ACD-MM Algorithm}
 \textbf{Input:} $\mathbf{\Pi},\eta,m,l$;\\
 \textbf{Output:} $\hat{f}(x)$;
 \begin{itemize}
 \item[\textbf{ 1:}]Determine metrics $\gamma_j$ as in (\ref{gamma}) and define $\Theta$;
 \item[\textbf{ 2:}]Formulate $2^{\eta}$ test-vectors $\underline{r}_u$ as in (\ref{tv});
 \item[\textbf{ 3:}]Transform all $\underline{r}_u$ into $\underline{z}_u$ as in (\ref{tv_transform});
 \item[\textbf{ 4:}]\textbf{For} each test-vector $\underline{z}_u$ \textbf{do}
 \item[\textbf{ 5:}]\quad Formulate $\tilde{\mathcal{B}}_l$ by (\ref{Ptu_1_map}) (\ref{Ptu_2_map}) and map to $\mathcal{A}_l$ by (\ref{mapping_re_encoding});
 \item[\textbf{ 6:}]\quad Reduce $\mathcal{A}_l$ into $\mathcal{A}_l'$ and demap it to $\tilde{\mathcal{B}}_l'$ by (\ref{demapping_re_encoding});
 \item[\textbf{ 7:}]\quad Construct $Q$ by (\ref{QfromM}) and (\ref{Q_restore});
 \item[\textbf{ 8:}]\quad Factorize $Q$ to find $\hat{f}(x)$;
 \item[\textbf{ 9:}]\textbf{End for}
\end{itemize}
\end{algorithm}

\section{The KV-MM Algorithm}
This section introduces the KV-MM algorithm. It transfers the reliability matrix into a multiplicity matrix that defines the MM interpolation. Its re-encoding transformed variant will also be introduced.

\subsection{From Reliability Matrix to Multiplicity Matrix}
The reliability matrix $\mathbf{\Pi}$ will be proportionally transformed into a multiplicity matrix $\mathbf{M}$ using Algorithm A of \cite{KV2003}, whose entry $m_{ij}$ indicates the interpolation multiplicity for point $(\alpha_j, \sigma_i)$. Hence, there are $|\{m_{ij}~|~m_{ij} \neq 0\}|$ interpolation points. Interpolation is to find the minimum polynomial $Q(x, y)$ that interpolates all points $(\alpha_j, \sigma_i)$ with a multiplicity of $m_{ij}$. Let $i_j=\text{index}\{\sigma_i~|~\sigma_i=c_j\}$, the codeword score is defined as $S_{\mathbf{M}}(\underline{c})=\sum_{j=0}^{n-1}m_{i_jj}$. What follows is a sufficient condition for a successful KV decoding.

\textit{\textbf{Theorem 3}} \cite{KV2003}. For an $(n,k)$ RS code, let $Q\in\mathbb{F}_q[x,y]$ denote an interpolated polynomial constructed based on $\mathbf{M}$. If $S_{\mathbf{M}}(\underline{c})>\deg_{1,k-1}Q(x,y)$, $Q(x,f(x))=0$.

The KV decoding is parameterized by the maximum decoding OLS $l$ and $l = \deg_yQ$. Given matrix $\mathbf{M}$, let us define
\begin{equation}
\mathsf{m}_j = \sum_{i = 0}^{q - 1}m_{ij}
\end{equation}
and
\begin{equation}
\mathsf{m} = \max\{\mathsf{m}_j,\forall j\}.
\end{equation}
The $\mathbf{\Pi}\rightarrow\mathbf{M}$ transform terminates when $\mathsf{m}=l$. The following subsection will show how to formulate module $\mathcal{M}_l$ based on $\mathbf{M}$.

\subsection{Module Formulation and Minimization}
We aim to generate module $\mathcal{M}_l$ whose bivariate polynomials interpolate points $(\alpha_j, \sigma_i)$ with a multiplicity of at least $m_{ij}$ and have a maximum $y$-degree of $l$. To generate $\mathcal{M}_l$, the following point enumeration is needed.

Let $L_j$ denote a list that enumerates interpolation points $(\alpha_j, \sigma_i)$ from column $j$ of $\mathbf{M}$ with their multiplicity $m_{ij}$ as
\begin{equation} \label{Lj}
L_j = [\underbrace{(\alpha_j, \sigma_i), \ldots, (\alpha_j, \sigma_i)}_{m_{ij}}, \forall i~\text{and}~m_{ij} \neq 0].
\end{equation}
Note that $|L_j| = \mathsf{m}_j$. Its balanced list $L_j'$ can be further created as follows. Copy one of the most frequent elements in $L_j$ to $L_j'$ and remove it from $L_j$. Repeat this process $\mathsf{m}_j$ times until $L_j$ becomes empty. Consequently, $L_j'$ can be denoted as
\begin{equation} \label{Ljpri}
L_j' = [(\alpha_j, y_j^{(0)}), (\alpha_j, y_j^{(1)}), \ldots, (\alpha_j, y_j^{(\mathsf{m}_j - 1)})],
\end{equation}
where $y_j^{(0)}, y_j^{(1)}, \ldots, y_j^{(\mathsf{m}_j - 1)} \in \mathbb{F}_q$ and they may not be distinct. $L_j'$ is a permutation of $L_j$ and $|L_j'| = \mathsf{m}_j$.
Let $\mathsf{m}_j(t)$ denote the maximum multiplicity of the last $\mathsf{m}_j - t$ elements of $L_j'$ as
\begin{equation} \label{mjs}
\mathsf{m}_j(t) = \max\{\text{multiplicity}((\alpha_j, y_j^{(\varepsilon)}))~|~\varepsilon = t, t + 1, \ldots, \mathsf{m}_j - 1\}.
\end{equation}
Note that $\mathsf{m}_j(0) = \max\{m_{ij},\forall i\}$ and $\mathsf{m}_j(\varepsilon) = 0$ for $\varepsilon \geq \mathsf{m}_j$.

The following example illustrates the above point enumeration process.

\begin{figure*}
  \centering
  \includegraphics[width=1.0\textwidth]{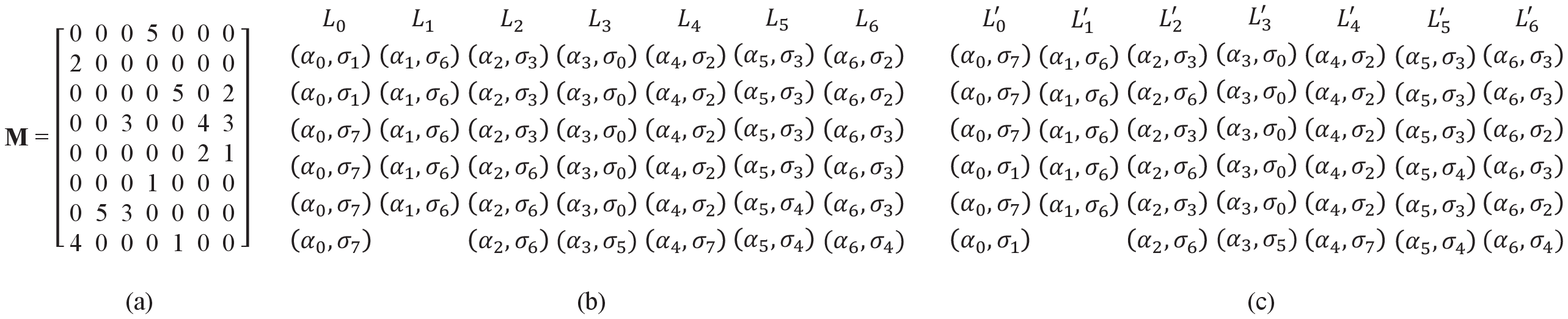}\\
  \caption{(a) Multiplicity matrix; (b) Enumeration lists $L_0 \sim L_6$; (c) Balanced lists $L_0' \sim L_6'$.}\label{num_lists}
\end{figure*}

\textit{\textbf{Example 1.}} In decoding a (7, 3) RS code, the multiplicity matrix is obtained as Fig. \ref{num_lists} (a). The enumeration lists $L_0 \sim L_6$ and their balanced lists $L_0' \sim L_6'$ are shown in Figs.~\ref{num_lists} (b) and \ref{num_lists} (c), respectively.
When $t = 1$, $\mathsf{m}_0(1) = 3$, $\mathsf{m}_1(1) = 4$, $\mathsf{m}_2(1) = 3$, $\mathsf{m}_3(1) = 4$, $\mathsf{m}_4(1) = 4$, $\mathsf{m}_5(1) = 3$ and $\mathsf{m}_6(1) = 2$.

%

Now, it is sufficient to formulate module $\mathcal{M}_l$. First, let us define
\begin{equation} \label{fvarx}
F_{\varepsilon}(x) = \sum_{j = 0}^{n - 1} y_j^{(\varepsilon)} \Phi_j(x),
\end{equation}
where $\varepsilon = 0, 1, \ldots, l - 1$. Based on (\ref{lagrange}), we have $F_{\varepsilon}(\alpha_j) = y_j^{(\varepsilon)},\forall j$. Hence, $y - F_{\varepsilon}(x)$ interpolates points $(\alpha_j, y_j^{(\varepsilon)})$ for all $j$. Note that in formulating polynomials $F_{0}(x) \sim F_{l - 1}(x)$, if $\mathsf{m}_j < l$, we assume $y_j^{(\varepsilon)} = 0$ for $\varepsilon \geq \mathsf{m}_j$. Now, $\mathcal{M}_l$ can be generated as an $\mathbb{F}_q[x]$-module by
\begin{equation} \label{Pt_KV}
P_t(x, y) = \prod_{j = 0}^{n - 1}(x - \alpha_j)^{\mathsf{m}_j(t)} \prod_{\varepsilon = 0}^{t - 1}(y - F_{\varepsilon}(x)),
\end{equation}
where $t = 0, 1, \ldots, l$. It can be seen that $\prod_{\varepsilon = 0}^{t - 1}(y - F_{\varepsilon}(x))$ interpolates the first $t$ points of all balanced lists, while $\prod_{j = 0}^{n - 1}(x - \alpha_j)^{\mathsf{m}_j(t)}$ interpolates the remaining points. Moreover, $\deg_y P_t(x,y)\leq l,\forall t$. Therefore, $P_t(x,y)\in\mathcal{M}_l$.

\textit{\textbf{Lemma 4}} \cite{Alekhnovich2005}. Let $\mathcal{Q}_t(x,y)=\sum_{\tau=0}^t \mathcal{Q}_t^{(\tau)}(x)y^{\tau}\in \mathcal{M}_l$ with $\deg_y \mathcal{Q}_t=t$, we have $\prod_{j = 0}^{n - 1}(x - \alpha_j)^{\mathsf{m}_j(t)}|\mathcal{Q}_t^{(t)}(x)$.

Consequently, we have the following Theorem.

\textit{\textbf{Theorem 5.}} Any element of $\mathcal{M}_l$ can be written as an $\mathbb{F}_q[x]$-linear combination of $P_t(x,y)$.

\begin{IEEEproof}
Assume that $\mathcal{Q}(x,y)=\sum_{\tau=0}^l\mathcal{Q}^{(\tau)}(x)y^{\tau}\in\mathcal{M}_l$ and let us write (\ref{Pt_KV}) as $P_t(x,y)=\sum_{\tau=0}^{t}P_t^{(\tau)}(x)y^{\tau}$. Since when $t=l$, $P_l^{(l)}(x)=1$, there exists a polynomial $p_l(x)\in\mathbb{F}_q[x]$ that enables $\mathcal{Q}_{l-1}(x,y)=\mathcal{Q}(x,y)-p_l(x)P_l(x,y)$
so that $\deg_y \mathcal{Q}_{l-1}=l-1$. Note that if $\deg_y \mathcal{Q}<l$, $p_l(x)=0$. Since $\mathcal{Q},P_l\in\mathcal{M}_l$, $\mathcal{Q}_{l-1}\in\mathcal{M}_l$. Continuing with $t=l-1$, $P_{l-1}^{(l-1)}(x)=\prod_{j = 0}^{n - 1}(x - \alpha_j)^{\mathsf{m}_j(l-1)}$. Based on Lemma 4, $\prod_{j = 0}^{n - 1}(x - \alpha_j)^{\mathsf{m}_j(l-1)}|\mathcal{Q}_{l-1}^{(l-1)}(x)$. Therefore, we can generate $\mathcal{Q}_{l-2}(x,y)$ by $\mathcal{Q}_{l-2}(x,y)=\mathcal{Q}_{l-1}(x,y)-p_{l-1}(x)P_{l-1}(x,y)$
so that $\deg_y \mathcal{Q}_{l-2}=l-2$. Following the above deduction until $t=0$, we have $P_{0}^{(0)}(x)=\prod_{j = 0}^{n - 1}(x - \alpha_j)^{\mathsf{m}_j(0)}$ and $\prod_{j = 0}^{n - 1}(x - \alpha_j)^{\mathsf{m}_j(0)}|\mathcal{Q}_{0}^{(0)}(x)$. Hence, there exists $p_0(x)$ that enables $\mathcal{Q}_{0}(x,y)-p_{0}(x)P_{0}(x,y)=0$.
Therefore, if $\mathcal{Q}\in\mathcal{M}_l$, it can be written as an $\mathbb{F}_q[x]$-linear combination of $P_t(x,y)$.
\end{IEEEproof}


The above Theorem reveals that (\ref{Pt_KV}) forms a basis $\mathcal{B}_l$ of $\mathcal{M}_l$. Presenting $\mathcal{B}_l$ as a matrix over $\mathbb{F}_q[x]$, we can generate $\mathcal{A}_l$ by (\ref{A_ml_1_map}). The MS algorithm will reduce $\mathcal{A}_l$ into $\mathcal{A}_l'$. Demap it by (\ref{D_ml_1_demap}). Polynomial $Q$ can be further retrieved.

The KV-MM algorithm is summarized as in Algorithm \ref{alg:The KV-MM Algorithm}.

\begin{algorithm}[htbp]
\caption{The KV-MM Algorithm}
\label{alg:The KV-MM Algorithm}
 \textbf{Input:} $\mathbf{M},l$;\\
 \textbf{Output:} $\hat{f}(x)$;
 \begin{itemize}
 \item[\textbf{ 1:}]Create balanced lists $L_0'\sim L_{n-1}'$;
 \item[\textbf{ 2:}]Formulate $\mathcal{B}_l$ by (\ref{Pt_KV}) and map to $\mathcal{A}_l$ by (\ref{A_ml_1_map});
 \item[\textbf{ 3:}]Reduce $\mathcal{A}_l$ into $\mathcal{A}_l'$ and demap it to $\mathcal{B}_l'$ by (\ref{D_ml_1_demap});
 \item[\textbf{ 4:}]Construct $Q$ by (\ref{QfromM});
 \item[\textbf{ 5:}]Factorize $Q$ to find $\hat{f}(x)$.
\end{itemize}
\end{algorithm}

\subsection{The Re-encoding Transformed KV-MM}
Similar to the ACD-MM algorithm, re-encoding transform helps reduce degree of the basis entries, reducing basis reduction complexity.
First, we will sort $\mathsf{m}_0(0),\mathsf{m}_1(0),\ldots,\mathsf{m}_{n-1}(0)$ to obtain an index sequence $j_0,j_1,\ldots,j_{n-1}$ such that $\mathsf{m}_{j_0}(0)\geq\mathsf{m}_{j_1}(0)\geq\cdots\geq\mathsf{m}_{j_{n-1}}(0)$. Let us define $\Upsilon=\{j_{0},j_{1},\ldots,j_{k-1}\}$ and $\bar{\Upsilon}=\{j_k,j_{k+1},\ldots,j_{n-1}\}$. The $k$ points $(\alpha_j,y_j^{(0)})$ where $j\in\Upsilon$ are chosen to form the re-encoding polynomial
\begin{equation}\label{H(x)}
H_{\Upsilon}(x)=\sum_{j\in\Upsilon}y_j^{(0)}\prod_{j'\in\Upsilon,j'\neq j}\frac{x-\alpha_{j'}}{\alpha_j-\alpha_{j'}},
\end{equation}
where $H_{\Upsilon}(\alpha_j)=y_j^{(0)},\forall j\in\Upsilon$. All interpolation points $(\alpha_j,y_j^{(\varepsilon)})$ will be transformed by
\begin{equation}\label{w_j^varepsilon}
(\alpha_j,w_j^{(\varepsilon)})=(\alpha_j,y_j^{(\varepsilon)}-H_{\Upsilon}(\alpha_j)).
\end{equation}
For $j\in\Upsilon$, if $y_j^{(\varepsilon)}=y_j^{(0)}$, $w_j^{(\varepsilon)}=0$. Let us define $\Lambda_\varepsilon=\{j~|~w_j^{(\varepsilon)}=0,j\in\Upsilon\}$ and $\bar{\Lambda}_\varepsilon=\Upsilon\backslash\Lambda_\varepsilon$. With the transformed interpolation points $(\alpha_j,w_j^{(\varepsilon)})$, $F_{\varepsilon}(x)$ of (\ref{fvarx}) can be redefined as
\begin{equation} \label{fvarx_refined}
F_{\varepsilon}(x) = \sum_{j = 0}^{n - 1} w_j^{(\varepsilon)} \Phi_j(x).
\end{equation}

Now let us define
\begin{equation}\label{phi(x)}
\phi(x)=\prod_{j\in\Upsilon}(x-\alpha_j)^{\mathsf{m}_j(0)}
\end{equation}
and
\begin{equation}\label{psi(x)}
\psi(x)=\prod_{j\in\Upsilon}(x-\alpha_j).
\end{equation}
The following Lemma characterizes module generators when re-encoding transform is applied.

\textit{\textbf{Lemma 6.}} Given the multiplicity matrix $\mathbf{M}$ and the transformed interpolation points of (\ref{w_j^varepsilon}), we have $\phi(x)|P_t(x,y\psi(x))$.
\begin{IEEEproof}
Due to its length, this proof is given in Appendix A.
\end{IEEEproof}

\begin{algorithm}[htbp]
\caption{The Re-encoding Transformed KV-MM Algorithm}
\label{alg:Re-encoding KV-MM Algorithm}
 \textbf{Input:} $\mathbf{M},l$;\\
 \textbf{Output:} $\hat{f}(x)$;
 \begin{itemize}
 \item[\textbf{ 1:}]Create balanced lists $L_0'\sim L_{n-1}'$;
 \item[\textbf{ 2:}]Sort $\mathsf{m}_0(0),\mathsf{m}_1(0),\ldots,\mathsf{m}_{n-1}(0)$ and define $\Upsilon$;
 \item[\textbf{ 3:}]Perform the re-encoding transform for all interpolation points as in (\ref{w_j^varepsilon});
 \item[\textbf{ 4:}]Formulate $\tilde{\mathcal{B}}_l$ by (\ref{KV_MM_re_encoding_generator}) and map to $\mathcal{A}_l$ by (\ref{mapping_re_encoding});
 \item[\textbf{ 5:}]Reduce $\mathcal{A}_l$ into $\mathcal{A}_l'$ and demap it to $\tilde{\mathcal{B}}_l'$ by (\ref{demapping_re_encoding});
 \item[\textbf{ 6:}]Construct $Q$ by (\ref{QfromM}) and (\ref{KV_restore});
 \item[\textbf{ 7:}]Factorize $Q$ to find $\hat{f}(x)$.
\end{itemize}
\end{algorithm}

We further define the following bijective mapping
\begin{equation}\label{Pt_KV_mapping}
\begin{aligned}
\varphi:~~~~ \mathcal{M}_l &\rightarrow \mathbb{F}_q[x,y]\\
P_t(x,y) &\mapsto \phi(x)^{-1}P_t(x,y\psi(x)).
\end{aligned}
\end{equation}
Consequently, polynomials of $\varphi(\mathcal{M}_l)$ would have a lower $x$-degree. This leads to a simpler basis reduction.
Let us further define
\begin{equation}
\tilde{w}_j^{(\varepsilon)}=\frac{w_j^{(\varepsilon)}}{\prod_{j'=0,j'\neq j}^{n-1}(\alpha_j-\alpha_{j'})}
\end{equation}
and
\begin{equation}
T_{\varepsilon}(x)=\sum_{j\in\bar\Upsilon\cup\bar\Lambda_\varepsilon}\tilde{w}_j^{(\varepsilon)}\prod_{j'\in\bar\Upsilon\cup\bar\Lambda_\varepsilon,j'\neq j}(x-\alpha_{j'}),
\end{equation}
the proof of Lemma 6 reveals that
\begin{equation}
P_t(x,y\psi(x))=\phi(x)\cdot\prod_{j\in\bar\Upsilon}(x-\alpha_j)^{\mathsf{m}_j(t)}\cdot\prod_{j\in\bar\Lambda_t}(x-\alpha_j)\cdot\prod_{\varepsilon=0}^{t-1}\Big(y\prod_{j\in\bar\Lambda_\varepsilon}(x-\alpha_j)-T_{\varepsilon}(x)\Big).
\end{equation}
Based on (\ref{Pt_KV_mapping}), $\varphi(\mathcal{M}_l)$ can be generated by
\begin{equation}\label{KV_MM_re_encoding_generator}
\tilde{P}_t(x,y)=\prod_{j\in\bar\Upsilon}(x-\alpha_j)^{\mathsf{m}_j(t)}\cdot\prod_{j\in\bar\Lambda_t}(x-\alpha_j)\cdot\prod_{\varepsilon=0}^{t-1}\Big(y\prod_{j\in\bar\Lambda_\varepsilon}(x-\alpha_j)-T_{\varepsilon}(x)\Big)
\end{equation}
and $t=0,1,\ldots,l$. Note that $\bar\Lambda_l=\emptyset$. The above $\tilde{P}_t(x,y)$ form a basis $\tilde{\mathcal{B}}_l$ of $\varphi(\mathcal{M}_l)$. Present $\tilde{\mathcal{B}}_l$ as a matrix over $\mathbb{F}_q[x]$ and perform the mapping of $\mathcal{A}_l=\tilde{\mathcal{B}}_l\cdot\tilde{\mathcal{D}}_{1, l}$.
The MS algorithm will reduce $\mathcal{A}_{l}$ into $\mathcal{A}_l'$. Demap it by $\tilde{\mathcal{B}}_l'=\mathcal{A}_l'\cdot\tilde{\mathcal{D}}_{-1, l}$ and
polynomial $\tilde{Q}$ can be further retrieved. Finally, the interpolated polynomial $Q$ can be constructed by
\begin{equation}\label{KV_restore}
Q(x,y)=\phi(x)\tilde{Q}\Big(x,\frac{y}{\psi(x)}\Big).
\end{equation}
If $f'(x)$ is a $y$-root of $Q$, $\hat{f}(x)$ is estimated by $\hat{f}(x)=f'(x)+H_{\Upsilon}(x)$.

The re-encoding transformed KV-MM algorithm is summarized as in Algorithm \ref{alg:Re-encoding KV-MM Algorithm}.

\section{Complexity Analysis}
This section analyzes complexity of the MM interpolation that consists of module formulation and minimization. It characterizes complexity of the ACD-MM and the KV-MM algorithms. The complexity refers to the number of finite field multiplications that is required to decode a codeword. Note that in the MM interpolation, finite field multiplication dominates.

\subsection{Without the Re-encoding Transform}
We first analyze complexity of the module formulation. For the ACD-MM algorithm, $G(x)$ and $\Phi_j(x)$ can be computed offline. Instead, we need $n^2$ multiplications to compute $R(x)$. The complexity of formulating $\mathcal{B}_l$ is $\sum_{t=0}^{m}\sum_{j=0}^{t}(t-j)(n-1)\cdot(m-t)n\approx\frac{1}{24}n^2(m+1)^4$. Note that we use the naive polynomial multiplication. Therefore, in the ACD-MM algorithm, complexity of the module formulation is $\frac{1}{24}n^2((m+1)^4+24)$ for each test-vector. With $\eta$ reliable symbols, the overall formulation complexity would be scaled up by $2^{\eta}$. For the KV-MM algorithm, the formulation of $F_{\varepsilon}(x)$ and $\mathcal{B}_l$ require $n^2l$ and $\frac{1}{24}n^2(l+1)^4$ multiplications, respectively. Its module formulation complexity is $\frac{1}{24}n^2((l+1)^4+24l)$.

Next we analyze complexity of the basis reduction. This is determined by the degree of $\mathcal{A}_l|_t^{(\tau)}$ and the number of row operations that reduce $\mathcal{A}_l$ into $\mathcal{A}_l'$.

\textit{\textbf{Lemma 7}} \cite{Nielsen2013phd}. Given a matrix $\mathcal{A}_{l}$ over $\mathbb{F}_q[x]$, there are less than $(l + 1)(\deg\mathcal{A}_{l} - \deg\det\mathcal{A}_{l} + l)$ row operations to reduce it into the weak Popov form $\mathcal{A}_l'$.

The following two Lemmas further characterize $\deg\mathcal{A}_l|_t^{(\tau)}$ and $\deg\mathcal{A}_l - \deg\det\mathcal{A}_l$.

\textit{\textbf{Lemma 8.}} Without the re-encoding transform, $\deg\mathcal{A}_l|_t^{(\tau)} \leq nl$.

\begin{IEEEproof}
For the ACD-MM algorithm, we can determine $\deg\mathcal{A}_l|_t^{(\tau)}$ based on the generators (\ref{Pt_1}) (\ref{Pt_2}) and mapping (\ref{A_ml_1_map}). When $0\leq t\leq m$,
\begin{equation*}
\begin{aligned}
\deg\mathcal{A}_l|_t^{(\tau)}&=n(m-t)+(n-1)(t-\tau)+(k-1)\tau\\
&=nm-t-(n-k)\tau.
\end{aligned}
\end{equation*}
Therefore, $\max\{\deg\mathcal{A}_l|_t^{(\tau)},0\leq t\leq m\}=\deg\mathcal{A}_l|_0^{(0)}=nm$. When $m<t\leq l$,
\begin{equation*}
\begin{aligned}
\deg\mathcal{A}_l|_t^{(\tau)}&=(n-1)(t-\tau)+(k-1)\tau\\
&=(n-1)t-(n-k)\tau,
\end{aligned}
\end{equation*}
and $\max\{\deg\mathcal{A}_l|_t^{(\tau)},m<t\leq l\}=\deg\mathcal{A}_l|_l^{(0)}=(n-1)l$. Hence, for the ACD-MM algorithm, $\max\{\deg\mathcal{A}_l|_t^{(\tau)},\forall (t,\tau)\}=\max\{nm,(n-1)l\}\leq nl$.

For the KV-MM algorithm, the generators (\ref{Pt_KV}) and mapping (\ref{A_ml_1_map}) lead to
\begin{equation*}
\begin{aligned}
\deg\mathcal{A}_l|_t^{(\tau)}&\leq n(l-t)+(n-1)(t-\tau)+(k-1)\tau\\
&=nl-t-(n-k)\tau.
\end{aligned}
\end{equation*}
Therefore, $\max\{\deg\mathcal{A}_l|_t^{(\tau)},\forall (t,\tau)\}=\deg\mathcal{A}_l|_0^{(0)}=nl$.
\end{IEEEproof}

\textit{\textbf{Lemma 9.}} Without the re-encoding transform, $\deg\mathcal{A}_l - \deg\det\mathcal{A}_l \leq \frac{1}{2}(n - k)(l^2 + l)$.

\begin{IEEEproof}
Due to its length, this proof is given in Appendix B.
\end{IEEEproof}

Based on Lemmas 7 to 9, we can conclude that the basis reduction process requires at most $\frac{1}{2}n(n - k)l^2(l + 1)^2(l+2)\approx\frac{1}{2}n(n - k)(l + 1)^5$ finite field multiplications.

Summarizing the above analysis, with $\eta$ reliable symbols, the ACD-MM algorithm requires at most $2^\eta(\frac{1}{24}n^2((m+1)^4+24)+\frac{1}{2}n(n - k)(l + 1)^5)$ finite field multiplications. With the maximum decoding OLS $l$, the KV-MM algorithm requires at most $\frac{1}{24}n^2((l+1)^4+24l)+\frac{1}{2}n(n - k)(l + 1)^5$ finite field multiplications. These conclusions reveal that the MM interpolation complexity would be smaller for high rate codes, which is of practical interest. Table \ref{Complexity of the MM interpolation of ACD and KV algorithms} shows our numerical results on the MM complexity. It verifies the above analysis.

\begin{table*}[htbp]
\centering
\caption{The MM interpolation complexity of the ACD and KV algorithms}
\label{Complexity of the MM interpolation of ACD and KV algorithms}
\begin{tabular}{ccccc}
\hline\hline
\multicolumn{2}{c}{}                                                                                     & (63, 31) RS       & (63, 47) RS      & (63, 55) RS      \\ \hline
\multicolumn{1}{c|}{\multirow{2}{*}{\begin{tabular}[c]{@{}c@{}}ACD-MM\end{tabular}}} & $(m,\eta)=(1,3)$  & $1.02\times10^5$  & $9.06\times10^4$ & $8.85\times10^4$ \\
\multicolumn{1}{c|}{}                                                                & $(m,\eta)=(5,3)$  & $3.34\times10^7$  & $-$  & $-$              \\ \hline
\multicolumn{1}{c|}{\multirow{2}{*}{\begin{tabular}[c]{@{}c@{}}KV-MM\end{tabular}}}  & $l = 4$           & $1.68\times10^6$  & $1.22\times10^6$ & $1.01\times10^6$ \\
\multicolumn{1}{c|}{}                                                                & $l = 8$           & $2.84\times10^7$  & $1.95\times10^7$ & $1.45\times10^7$ \\ \hline\hline
\end{tabular}
\end{table*}

%
%
%

\subsection{With the Re-encoding Transform}
Re-encoding transform reduces the degree of module generators, leading to a simpler basis reduction. We first look into the complexity of formulating the re-encoding polynomial and the module basis $\tilde{\mathcal{B}}_l$. To compute $H_{\Theta}(x)$ (or $H_{\Upsilon}(x)$), we need $k(\frac{(k-1)k}{2}+(k-1)+k)\approx\frac{1}{2}k(k+1)^2$ finite field multiplications. The following interpolation point transform requires $(n-k)(k-1)$ multiplications. In formulating $\tilde{\mathcal{B}}_l$ for the ACD-MM algorithm, computing $\tilde{G}(x)$ and $\tilde{R}_u(x)$ require $\frac{1}{2}(n-k)^2$ and $\frac{1}{2}(n-k)^2(n-k-1)+(n-k)(n-k-1)+n(n-k)\approx\frac{1}{2}(n-k)((n-k)^2+2n)$ multiplications, respectively. Further formulating $\tilde{\mathcal{B}}_l$ requires $\sum_{t=0}^{m}\sum_{j=0}^{t}(t-j)(n-k-1)\cdot(m-t)(n-k)\approx\frac{1}{24}(n-k)^2(m+1)^4$ multiplications. Therefore, the formulation complexity for each test-vector is $\frac{1}{24}(n-k)^2((m+1)^4+12(n-k))$.
For the KV-MM algorithm, computing $T_{\varepsilon}(x)$ and $\tilde{\mathcal{B}}_l$ require $\frac{1}{2}(n-k)((n-k)^2+2n)$ and $\frac{1}{24}(n-k)^2(l+1)^4$ finite field multiplications \footnote{It is assumed that $\bar{\Lambda}_{\varepsilon}=\emptyset,\forall \varepsilon$.}, respectively. Therefore, complexity of this formulation is $\frac{1}{24}(n-k)^2((l+1)^4+12(n-k))$.

Next, we analyze complexity of the basis reduction.

\textit{\textbf{Lemma 10.}} With the re-encoding transform, $\deg\mathcal{A}_l|_t^{(\tau)} \leq (n - k + 1)l$.

\begin{IEEEproof}
For the ACD-MM algorithm, when $0\leq t\leq m$,
\begin{equation*}
\begin{aligned}
\deg\mathcal{A}_l|_t^{(\tau)}&=(n-k)(m-t)+(n-k-1)(t-\tau)+(l-\tau).\\
\end{aligned}
\end{equation*}
Since $m\leq l$, $\max\{\deg\mathcal{A}_l|_t^{(\tau)},0\leq t\leq m\}=\deg\mathcal{A}_l|_0^{(0)}=(n-k+1)l$.
When $m<t\leq l$,
\begin{equation*}
\begin{aligned}
\deg\mathcal{A}_l|_t^{(\tau)}&=k(t-m)+(n-k-1)(t-\tau)+(l-\tau),
\end{aligned}
\end{equation*}
and $\max\{\deg\mathcal{A}_l|_t^{(\tau)},m< t\leq l\}=\deg\mathcal{A}_l|_l^{(0)}=(n-k)l$.

With the re-encoding transform, entry size of $\mathcal{A}_l$ in the KV-MM algorithm is
\begin{equation*}
\begin{aligned}
\deg\mathcal{A}_l|_t^{(\tau)}&\leq (n-k)(l-t)+(n-k-1)(t-\tau)+(l-\tau).\\
\end{aligned}
\end{equation*}
Therefore, $\max\{\deg\mathcal{A}_l|_t^{(\tau)},\forall(t,\tau)\}=\deg\mathcal{A}_l|_0^{(0)}=(n-k+1)l$.
\end{IEEEproof}

\textit{\textbf{Lemma 11.}} With the re-encoding transform, $\deg\mathcal{A}_l - \deg\det\mathcal{A}_l \leq \frac{1}{2}(n - k)(l^2 + l)$.

\begin{IEEEproof}
For the ACD-MM algorithm, we have
\begin{equation*}
\deg\mathcal{A}_l=\sum_{t=0}^{m}((n-k)m-t+l)+\sum_{t=m+1}^{l}((n-1)t+(l-km))
\end{equation*}
and
\begin{equation*}
\deg\det\mathcal{A}_l=\sum_{t=0}^{m}((n-k)(m-t)+(l-t))+\sum_{t=m+1}^{l}(k(t-m)+(l-t)).
\end{equation*}
Hence, $\deg\mathcal{A}_l-\deg\det\mathcal{A}_l=\sum_{t=0}^{l}(n-k)t=\frac{1}{2}(n-k)(l^2+l)$.

Similar to the proof of Lemma 9, let $\tau_t$ identify the maximum entry of $\mathcal{A}_l|_t$. For the KV-MM algorithm, we have
\begin{equation*}
\begin{aligned}
&\deg\mathcal{A}_l-\deg\det\mathcal{A}_l
\leq&\sum_{t=0}^{l}((n-k+1)(t-\tau_t)+(l-\tau_t)-(l-t)).\\
\end{aligned}
\end{equation*}
When $\tau_t=0$, $\max\{\deg\mathcal{A}_l-\deg\det\mathcal{A}_l\}=\frac{1}{2}(n-k)(l^2+l)$.
\end{IEEEproof}

Therefore, despite the re-encoding transform can reduce degree of basis entries, it does not attribute to reducing the number of row operations during the basis reduction process.

Based on Lemmas 7, 10 and 11, we know that with the re-encoding transform, the basis reduction requires at most $\frac{1}{2}(n-k)^2(l+1)^5$ finite field multiplications. Therefore, for the re-encoding transformed variants, the MM interpolation requires $2^\eta(\frac{1}{24}(n-k)^2((m+1)^4+12(n-k))+\frac{1}{2}(n-k)^2(l+1)^5)$ and $\frac{1}{24}(n-k)^2((l+1)^4+12(n-k))+\frac{1}{2}(n-k)^2(l+1)^5$ multiplications in the ACD-MM and the KV-MM algorithms, respectively.

Compared with the earlier analysis, re-encoding transform helps reduce the MM complexity by a factor of $\frac{k}{n}$. Again, this result favors high rate codes. It can be noticed that
when $m$ (or $l$) is small, the complexity reduction brought by the re-encoding transform may not compensate the extra re-encoding computation. But as $m$ (or $l$) increases, the re-encoding complexity reduction effect would emerge.
Table \ref{Complexity of the MM interpolation of re-encoding transformed ACD and KV algorithms} shows our numerical results on complexity of the MM interpolation when the re-encoding transform is applied. Again, these results verify the above analysis.

%
%

\begin{table*}[htbp]
\centering
\caption{The MM interpolation complexity of the re-encoding transformed ACD and KV algorithms}
\label{Complexity of the MM interpolation of re-encoding transformed ACD and KV algorithms}
\begin{tabular}{ccccc}
\hline\hline
\multicolumn{2}{c}{}                                                                                     & (63, 31) RS       & (63, 47) RS      & (63, 55) RS      \\ \hline
\multicolumn{1}{c|}{\multirow{2}{*}{\begin{tabular}[c]{@{}c@{}}ACD-MM\end{tabular}}} & $(m,\eta)=(1,3)$  & $1.65\times10^6$  & $1.37\times10^6$ & $1.21\times10^6$ \\
\multicolumn{1}{c|}{}                                                                & $(m,\eta)=(5,3)$  & $2.11\times10^7$  &  $-$  & $-$              \\ \hline
\multicolumn{1}{c|}{\multirow{2}{*}{\begin{tabular}[c]{@{}c@{}}KV-MM\end{tabular}}}  & $l = 4$           & $1.35\times10^6$  & $5.83\times10^5$ & $4.47\times10^5$ \\
\multicolumn{1}{c|}{}                                                                & $l = 8$           & $9.87\times10^6$  & $3.79\times10^6$ & $2.97\times10^6$ \\ \hline\hline
\end{tabular}
\end{table*}

\section{Simulation Results}
This section presents simulation results of the ACD and the KV algorithms. They are obtained over the additive white Gaussian noise (AWGN) channel using BPSK modulation. We aim to show their competency in terms of decoding performance and complexity, giving more insights of the potential applications. In this section, the decoding complexity is measured as the average number of finite field arithmetic operations that is required to decode a codeword, including the root-finding and the re-encoding transform. We will also show the MM interpolation's complexity advantage over Koetter's interpolation. We denote the ACD and the KV algorithms that employ Koetter's interpolation as the ACD-Koetter and the KV-Koetter algorithms, respectively. Note that the ACD-Koetter algorithm is the so called LCC algorithm \cite{BK_LCC2010}.

\subsection{Decoding Performance}

\begin{figure}[htbp]
  \centering
  \includegraphics[width=0.75\textwidth]{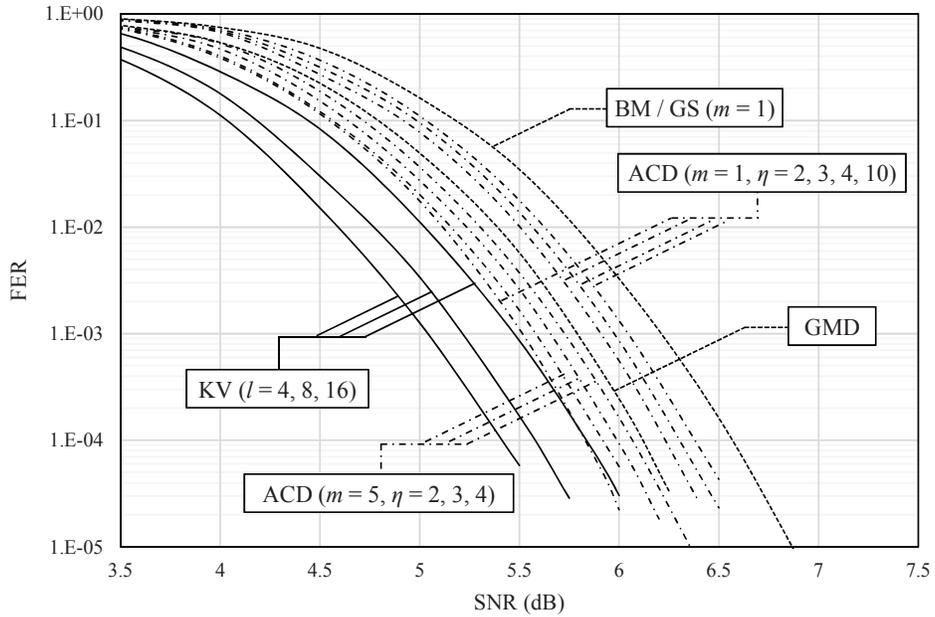}\\
  \caption{Performance of the (63, 31) RS code.}\label{(63,31)ACD_KV_MM_performance}
\end{figure}

Fig. \ref{(63,31)ACD_KV_MM_performance} shows the ACD and KV performance in decoding the (63, 31) RS code. Both of the ASD algorithms outperform the BM algorithm thanks to their soft decoding feature. When $m=1$, the ACD algorithm cannot outperform the GMD algorithm. But when $m=5$, each Chase decoding trial can correct at most 18 symbol errors. As a result, the ACD algorithm can outperform the GMD algorithm. Note that when $m=1$, the ACD algorithm with $\eta=2,3,4$ yields the same maximum decoding OLS as the KV algorithm with $l=4,8,16,$ respectively. Under such a benchmark, our results show the KV algorithm is more competent in error correction.

\begin{figure}[htbp]
  \centering
  \includegraphics[width=0.75\textwidth]{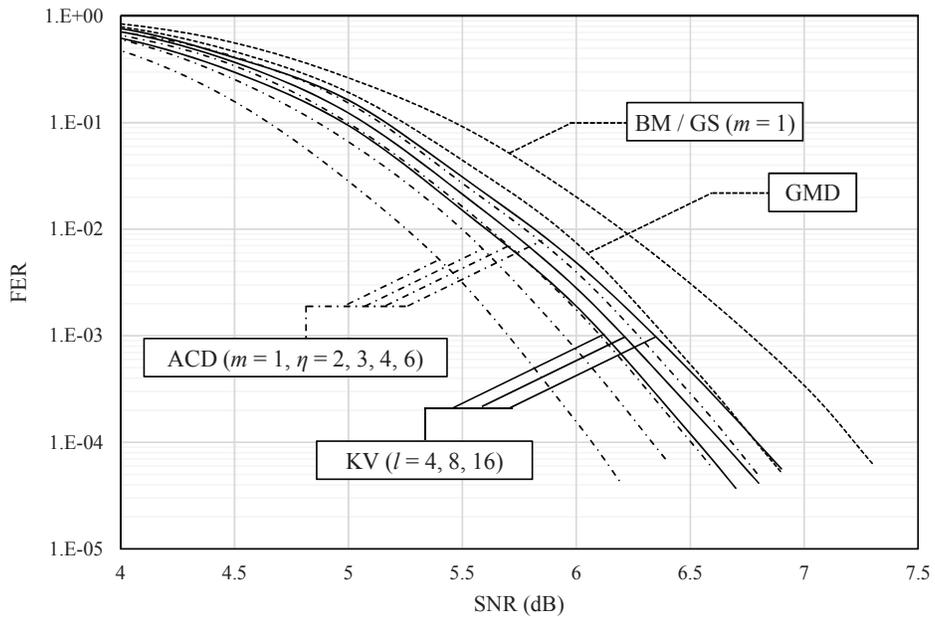}\\
  \caption{Performance of the (63, 55) RS code.}\label{(63,55)ACD_MM_performance}
\end{figure}

\begin{figure}[htbp]
  \centering
  \includegraphics[width=0.75\textwidth]{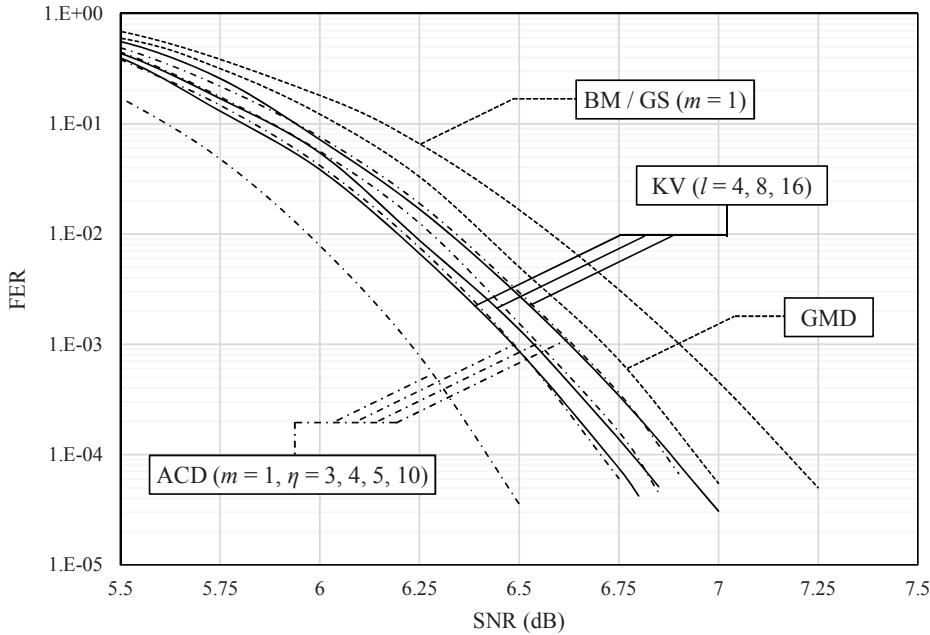}\\
  \caption{Performance of the (255, 239) RS code.}\label{(255,239)ACD_KV_MM_performance}
\end{figure}

Fig. \ref{(63,55)ACD_MM_performance} compares the two ASD algorithms in decoding the (63, 55) RS code. For this high rate code, both of the ASD algorithms outperform the BM and the GMD algorithms. In contrast to the (63, 31) RS code, the ACD algorithm outperforms the KV algorithm with the same maximum decoding OLS. For the ACD algorithm, more significant performance gain can be achieved by increasing its decoding parameter $\eta$. This is opposite in the KV algorithm. The KV algorithm is more effective for codes with a low to medium rate.
Fig. \ref{(255,239)ACD_KV_MM_performance} further shows performance of the popular (255, 239) RS code. With the same maximum decoding OLS, e.g., the KV $(l=8)$ and the ACD $(\eta=3)$, and the KV $(l=16)$ and the ACD $(\eta=4)$, the KV algorithm outperforms the ACD algorithm. This is because the codebook cardinality of (255, 239) RS code is large, the ACD algorithm needs a larger $\eta$ to improve its decoding performance.

\subsection{Complexity Comparison}

\begin{table}[htbp]
\centering
\caption{Complexity of the ACD algorithm in decoding the (63, 31) RS code}
\label{ACD complexity of (63, 31)}
\begin{tabular}{ccccc}
\hline\hline
\multirow{2}{*}{$(m,\eta)$} & \multicolumn{2}{c}{w/o re-encoding} & \multicolumn{2}{c}{with re-encoding} \\
                            & ACD-MM           & ACD-Koetter      & ACD-MM            & ACD-Koetter      \\ \hline
(1, 3) & $2.96\times10^5$  & $3.17\times10^5$     & $1.74\times10^6$   & $1.71\times10^5$    \\

(5, 3) & $5.37\times10^7$  & $1.71\times10^8$     & $2.28\times10^7$   & $5.73\times10^7$    \\

(1, 10)& $3.81\times10^7$  & $2.46\times10^7$     & $1.89\times10^8$   & $1.16\times10^7$    \\

\hline\hline
\end{tabular}
\end{table}

%

\begin{table}[htbp]
\centering
\caption{Complexity of the KV algorithm in decoding the (63, 31) RS code}
\label{KV complexity of (63, 31)}
\begin{tabular}{ccccc}
\hline\hline
\multirow{2}{*}{$l$} & \multicolumn{2}{c}{w/o re-encoding} & \multicolumn{2}{c}{with re-encoding} \\
  & KV-MM           & KV-Koetter      & KV-MM            & KV-Koetter      \\ \hline
4 & $1.82\times10^6$  & $1.59\times10^7$     & $1.48\times10^6$   & $6.16\times10^6$    \\
8 & $3.01\times10^7$  & $3.50\times10^8$     & $1.11\times10^7$   & $1.10\times10^8$    \\ \hline\hline
\end{tabular}
\end{table}

Tables \ref{ACD complexity of (63, 31)} and \ref{KV complexity of (63, 31)} show the ACD and KV complexity in decoding the (63, 31) RS code, respectively. We first compare complexity of the ACD-MM and ACD-Koetter algorithms, as well as that of the KV-MM and KV-Koetter algorithms. For the ACD algorithm, when $m$ is large, the MM interpolation yields a lower complexity than Koetter's interpolation despite whether the re-encoding transform applied. However, when $m=1$, using the MM interpolation may not be less complex. This is because using Koetter's interpolation, Gr\"obner basis of all test-vectors can be generated in a binary tree growth fashion, forging a low complexity \cite{BK_LCC2010}. Moreover, when $m=1$, the re-encoding transformed ACD-MM algorithm becomes more complex. This extra computation comes from the re-encoding computation. The complexity reduction brought by re-encoding cannot compensate this extra computation. However, as $m$ increases, the complexity reduction effect of re-encoding will emerge. For the KV algorithm, MM interpolation yields a lower complexity than Koetter's interpolation, despite whether the re-encoding transform is applied. The above results also show the re-encoding transform yields a complexity reduction factor of $\frac{k}{n}$. This verifies the earlier analysis in Section V. Fig. \ref{(63,31)ACD_KV_MM_performance} shows that the performance of KV $(l=4)$ is similar to that of the ACD $(1,10)$. Tables \ref{ACD complexity of (63, 31)} and \ref{KV complexity of (63, 31)} show that the KV algorithm is less complex than the ACD algorithm. On the other hand, with a similar decoding complexity, e.g., the re-encoding transformed ACD-Koetter $(1,10)$ and the KV-MM $(l=8)$, the KV algorithm outperforms.

%
%

\begin{table}[htbp]
\centering
\caption{Complexity of the re-encoding transformed ACD and KV algorithms in decoding the (63, 55) RS code}
\label{KV complexity of (63, 55)}
\begin{tabular}{cccccc}
\hline\hline
$(m,\eta)$ & ACD-MM & ACD-Koetter & $l$ & KV-MM & KV-Koetter \\ \hline
(1, 2) & $1.14\times10^6$  & $3.58\times10^5$  & 8    & $4.10\times10^6$   & $2.39\times10^7$    \\

(1, 3) & $1.36\times10^6$  & $4.27\times10^5$  & 16   & $3.60\times10^7$   & $2.86\times10^8$    \\

(1, 6) & $4.43\times10^6$  & $1.26\times10^6$  &  $-$   & $-$   & $-$    \\
\hline\hline
\end{tabular}
\end{table}

Table \ref{KV complexity of (63, 55)} shows the decoding complexity of the re-encoding transformed ACD and KV algorithms for the (63, 55) RS code. Fig. \ref{(63,55)ACD_MM_performance} shows that the ACD $(\eta=2)$ and $(\eta=3)$ perform similarly as the KV $(l=8)$ and $(l=16)$, respectively. Table \ref{KV complexity of (63, 55)} shows that the ACD algorithm is less complex. With a similar decoding complexity, e.g., the ACD-MM $(\eta=6)$ and the KV-MM $(l=8)$, the ACD-MM algorithm prevails in performance. Hence, for high rate codes, the ACD algorithm becomes more competent in error correction.

\begin{table}[htbp]
\centering
\caption{Complexity of the re-encoding transformed ACD and KV algorithms in decoding the (255, 239) RS code}
\label{KV_ACD_complexity of (255, 239)}
\begin{tabular}{cccccc}
\hline\hline
$(m,\eta)$   & ACD-MM           & ACD-Koetter        & $l$ & KV-MM            & KV-Koetter              \\ \hline
(1, 3)       & $8.35\times10^7$ & $2.43\times10^7$   &4    & $1.63\times10^7$ & $1.06\times10^8$        \\
(1, 4)       & $9.51\times10^7$ & $2.71\times10^7$   &8    & $1.27\times10^8$ & $3.24\times10^8$        \\
(1, 10)      & $1.56\times10^9$ & $2.80\times10^8$   &16   & $9.51\times10^8$ & $2.51\times10^9$        \\ \hline\hline
\end{tabular}
\end{table}

%
%

Finally, Table \ref{KV_ACD_complexity of (255, 239)} further shows complexity of the re-encoding transformed ACD and KV algorithms in decoding the (255, 239) RS code. It shows that for the KV algorithm, using MM interpolation incurs a lower complexity. But this is not the case for the ACD algorithm. Fig. \ref{(255,239)ACD_KV_MM_performance} also shows that the ACD $(\eta=4)$ performs similarly as the KV $(l=8)$. Table \ref{KV_ACD_complexity of (255, 239)} shows the ACD algorithm is less complex using either interpolation technique. With a similar decoding complexity, e.g., the ACD-MM $(\eta=10)$ and the KV-MM $(l=16)$, the ACD-MM algorithm prevails in performance.

\section{Conclusions}
This paper has presented the low-complexity ACD-MM and KV-MM algorithms for RS codes. Unlike Koetter's interpolation that generates the Gr\"obner basis in a point-by-point fashion, the MM interpolation formulates a basis of module and further reduces it into the Gr\"obner basis. It can yield a smaller computational cost. Re-encoding transformed variants of the two ASD algorithms have also been presented. They have a simpler basis reduction process since re-encoding transform helps reduce the degree of module generators. Our complexity analysis has shown that both the MM interpolation and the re-encoding transform are more effective in yielding a low complexity for high rate codes. These results fall into the interest of practice in which high rate codes are widely employed. Our simulation results have verified that the MM interpolation enables a lighter decoding computation for the two ASD algorithms, despite whether the re-encoding transform is applied. A comprehensive decoding performance and complexity comparison between the two ASD algorithms has also been presented. For medium rate codes, the KV algorithm outperforms the ACD algorithm under a similar complexity. While for high rate codes, the ACD algorithm prevails.

\vspace{1.5cm}
\section*{Appendix A\\Proof of Lemma 6}
With the module generators defined by (\ref{Pt_KV}), $P_t(x,y\psi(x))$ can be written as
\begin{equation*}\label{Pt(x,y*psi(x))}
P_t(x, y\psi(x)) = \prod_{j = 0}^{n - 1}(x - \alpha_j)^{\mathsf{m}_j(t)} \prod_{\varepsilon = 0}^{t - 1}(y\psi(x) - F_{\varepsilon}(x)).
\end{equation*}
Based on (\ref{phi(x)}), we know
\begin{flalign}
\prod_{j=0}^{n-1}(x-\alpha_j)^{\mathsf{m}_j(t)}&=\prod_{j\in\bar\Upsilon}(x-\alpha_j)^{\mathsf{m}_j(t)}\prod_{j\in\Upsilon}(x-\alpha_j)^{\mathsf{m}_j(t)} \nonumber\\
&=\phi(x)\cdot\prod_{j\in\bar{\Upsilon}}(x-\alpha_j)^{\mathsf{m}_j(t)}\prod_{j\in\Upsilon}(x-\alpha_j)^{\mathsf{m}_j(t)-\mathsf{m}_j(0)}.\label{first_part}
\end{flalign}

Based on the $F_{\varepsilon}(x)$ of (\ref{fvarx_refined}) and the fact that $w_j^{(\varepsilon)}=0,\forall j\in\Lambda_\varepsilon$, we have
\begin{equation*}
\begin{aligned}
&y\psi(x)-F_{\varepsilon}(x)\\
=&y\psi(x)-\sum_{j=0}^{n-1}w_j^{(\varepsilon)}\prod_{j'=0,j'\neq j}^{n-1}\frac{x-\alpha_{j'}}{\alpha_j-\alpha_{j'}}\\
=&y\psi(x)-\sum_{j\in\bar\Upsilon\cup\bar\Lambda_\varepsilon}w_j^{(\varepsilon)}\prod_{j'=0,j'\neq j}^{n-1}\frac{x-\alpha_{j'}}{\alpha_j-\alpha_{j'}}.
\end{aligned}
\end{equation*}
Let us denote
\begin{equation*}
\tilde{w}_j^{(\varepsilon)}=\frac{w_j^{(\varepsilon)}}{\prod_{j'=0,j'\neq j}^{n-1}(\alpha_j-\alpha_{j'})},
\end{equation*}
then
\begin{flalign}
&y\psi(x)-F_{\varepsilon}(x)\nonumber\\
=&y\psi(x)-\sum_{j\in\bar\Upsilon\cup\bar\Lambda_\varepsilon}\tilde{w}_j^{(\varepsilon)}\prod_{j'=0,j'\neq j}^{n-1}(x-\alpha_{j'}) \nonumber\\
=&y\psi(x)-\prod_{j'\in\Lambda_\varepsilon}(x-\alpha_{j'})\sum_{j\in\bar\Upsilon\cup\bar\Lambda_\varepsilon}\tilde{w}_j^{(\varepsilon)}\prod_{j'\in\bar\Upsilon\cup\bar\Lambda_\varepsilon,j'\neq j}(x-\alpha_{j'}) \nonumber\\
=&y\psi(x)-\prod_{j'\in\Lambda_\varepsilon}(x-\alpha_{j'})\cdot T_{\varepsilon}(x),\nonumber
\end{flalign}
where
\begin{equation*}
T_{\varepsilon}(x)=\sum_{j\in\bar\Upsilon\cup\bar\Lambda_\varepsilon}\tilde{w}_j^{(\varepsilon)}\prod_{j'\in\bar\Upsilon\cup\bar\Lambda_\varepsilon,j'\neq j}(x-\alpha_{j'}).
\end{equation*}
Based on (\ref{psi(x)}), we know $\psi(x)=\prod_{j\in\Lambda_\varepsilon}(x-\alpha_j)\prod_{j\in\bar\Lambda_\varepsilon}(x-\alpha_j)$,
\begin{equation*}
\begin{aligned}
&y\psi(x)-\prod_{j\in\Lambda_\varepsilon}(x-\alpha_{j})\cdot T_{\varepsilon}(x)
=&\prod_{j\in\Lambda_\varepsilon}(x-\alpha_j)\cdot\Big(y\prod_{j\in\bar\Lambda_\varepsilon}(x-\alpha_j)-T_{\varepsilon}(x)\Big).
\end{aligned}
\end{equation*}
Therefore,
\begin{equation}\label{second_part_rewritten}
\prod_{\varepsilon=0}^{t-1}(y\psi(x)-F_{\varepsilon}(x))=\prod_{\varepsilon=0}^{t-1}\prod_{j\in\Lambda_\varepsilon}(x-\alpha_j)\cdot\prod_{\varepsilon=0}^{t-1}\Big(y\prod_{j\in\bar\Lambda_\varepsilon}(x-\alpha_j)-T_{\varepsilon}(x)\Big).
\end{equation}

We now derive an equivalent expression for $\prod_{\varepsilon=0}^{t-1}\prod_{j\in\Lambda_\varepsilon}(x-\alpha_j)$. Let us first define the transformed balanced list $\mathcal{L}_j$ as
\begin{equation*}
\mathcal{L}_j=[(\alpha_j,\tilde{w}_j^{(0)}),(\alpha_j,\tilde{w}_j^{(1)}),\ldots,(\alpha_j,\tilde{w}_j^{(\mathsf{m}_j-1)})],
\end{equation*}
where $\tilde{w}_j^{(\varepsilon)}=\frac{y_j^{(\varepsilon)}-H_{\Upsilon}(\alpha_j)}{\prod_{j'=0,j'\neq j}^{n-1}(\alpha_j-\alpha_{j'})}$. Partition it into two disjoint sets as $\mathcal{L}_j(t)=[(\alpha_j,\tilde{w}_j^{(0)})$, $(\alpha_j,\tilde{w}_j^{(1)})$, $\ldots$, $(\alpha_j,\tilde{w}_j^{(t-1)})]$ and $\bar{\mathcal{L}}_j(t)=[(\alpha_j,\tilde{w}_j^{(t)}),(\alpha_j,\tilde{w}_j^{(t+1)}),\ldots,(\alpha_j,\tilde{w}_j^{(\mathsf{m}_j-1)})]$.
Let $\theta_j(t)$ denote the multiplicity of $(\alpha_j,\tilde{w}_j^{(0)})$ in $\mathcal{L}_j(t)$. Since $w_j^{(0)}=0$, $\forall j\in\Upsilon$, $\tilde{w}_j^{(0)}=0$ and $\theta_j(t)$ is the multiplicity of $(\alpha_j,0)$ of $\mathcal{L}_j(t)$ where $j\in\Upsilon$. Since $\Lambda_\varepsilon=\{j~|~w_j^{(\varepsilon)}=0,j\in\Upsilon\}$, we have
\begin{equation}\label{theta_j(t)}
\begin{aligned}
\prod_{\varepsilon=0}^{t-1}\prod_{j\in\Lambda_\varepsilon}(x-\alpha_j)=\prod_{j\in\Upsilon}(x-\alpha_j)^{\theta_j(t)}.
\end{aligned}
\end{equation}

Based on (\ref{first_part}), (\ref{second_part_rewritten}) and (\ref{theta_j(t)}), we have
\begin{equation*}\label{P_t(x,y)_re-written}
P_t(x,y\psi(x))=\phi(x)\cdot\prod_{j\in\bar\Upsilon}(x-\alpha_j)^{\mathsf{m}_j(t)}\cdot U_t(x)\cdot\prod_{\varepsilon=0}^{t-1}\Big(y\prod_{j\in\bar\Lambda_\varepsilon}(x-\alpha_j)-T_{\varepsilon}(x)\Big),
\end{equation*}
where $U_t(x)=\prod_{j\in\Upsilon}(x-\alpha_j)^{\mathsf{m}_j(t)-\mathsf{m}_j(0)+\theta_j(t)}$.
$U_t(x)$ can be simplified as follows. Let $\chi_j(t)$ denote the multiplicity of $(\alpha_j,0)$ in $\bar{\mathcal{L}}_j(t)$, we have $\mathsf{m}_j(0)=\theta_j(t)+\chi_j(t),\forall j\in\Upsilon$.
For $j\in\Upsilon$, if $\tilde{w}_j^{(t)}=\tilde{w}_j^{(0)}=0$, $\chi_j(t)=\mathsf{m}_j(t)$ and $\mathsf{m}_j(0)=\theta_j(t)+\mathsf{m}_j(t)$. Otherwise, if $\tilde{w}_j^{(t)}\neq0$, $\chi_j(t)=\mathsf{m}_j(t)-1$ and $\mathsf{m}_j(0)=\theta_j(t)+\mathsf{m}_j(t)-1$. Hence, when $j\in\Lambda_t$, $\mathsf{m}_j(t)-\mathsf{m}_j(0)+\theta_j(t)=0$. When $j\in\bar\Lambda_t$, $\mathsf{m}_j(t)-\mathsf{m}_j(0)+\theta_j(t)=1$. Therefore, $U_t(x)=\prod_{j\in\bar\Lambda_t}(x-\alpha_j)$.
As a result,
\begin{equation*}
P_t(x,y\psi(x))=\phi(x)\cdot\prod_{j\in\bar\Upsilon}(x-\alpha_j)^{\mathsf{m}_j(t)}\cdot\prod_{j\in\bar\Lambda_t}(x-\alpha_j)\cdot\prod_{\varepsilon=0}^{t-1}\Big(y\prod_{j\in\bar\Lambda_\varepsilon}(x-\alpha_j)-T_{\varepsilon}(x)\Big).
\end{equation*}
\hfill$\blacksquare$

\section*{Appendix B\\Proof of Lemma 9}
For the ACD-MM algorithm, when $0\leq t\leq m$, $\deg\mathcal{A}_l|_t=nm-t$. When $m<t\leq l$, $\deg\mathcal{A}_l|_t=(n-1)t$. Hence, we have
\begin{equation*}
\begin{aligned}
\deg\mathcal{A}_l&=\sum_{t=0}^{m}(nm-t)+\sum_{t=m+1}^{l}(n-1)t\\
\end{aligned}
\end{equation*}
and
\begin{equation*}
\begin{aligned}
\deg\det\mathcal{A}_{l}
&=\sum_{t=0}^{m}n(m-t)+\sum_{t=0}^{l}(k-1)t.\\
\end{aligned}
\end{equation*}
Therefore, $\deg\mathcal{A}_{l}-\deg\det\mathcal{A}_{l}=\sum_{t=0}^{l}(n-k)t=\frac{1}{2}(n-k)(l^2+l)$.

For the KV-MM algorithm, its generator of (\ref{Pt_KV}) can be written as $P_t(x,y)=\mathcal{G}_t(x)\cdot \mathcal{W}_t(x,y)$,
where $\mathcal{G}_t(x)=\prod_{j=0}^{n-1}(x-\alpha_j)^{\mathsf{m}_j(t)}$ and $\mathcal{W}_t(x,y)=\prod_{\varepsilon=0}^{t-1}(y-F_{\varepsilon}(x))=\sum_{\tau=0}^t \mathsf{w}_t^{(\tau)}(x)y^{\tau}$. Further based on mapping (\ref{A_ml_1_map}), we have $\mathcal{A}_l|_t^{(\tau)}=\mathcal{G}_t(x)\cdot \mathsf{w}_t^{(\tau)}(x)\cdot x^{(k-1)\tau}$. Let $\tau_t=\arg\max\{\deg \mathcal{A}_l|_t^{(\tau)},\forall \tau\}$ identify the maximum entry of $\mathcal{A}_l|_t$ such that $\deg\mathcal{A}_l|_t^{(\tau_t)}\geq\deg\mathcal{A}_l|_t^{(\tau)},\forall\tau\neq\tau_t$. Therefore, $\deg\mathcal{A}_{l}|_t=\deg (\mathcal{G}_t(x)\cdot \mathsf{w}_t^{(\tau_t)}(x)\cdot x^{(k-1)\tau_t})$ and
\begin{equation*}
\begin{aligned}
\deg\mathcal{A}_l
&=\sum_{t=0}^{l}(\deg \mathcal{G}_t(x)+\deg \mathsf{w}_t^{(\tau_t)}(x)+(k-1)\tau_t).
\end{aligned}
\end{equation*}
Since $\mathcal{A}_l$ is a lower-triangle matrix and $\mathsf{w}_t^{(t)}(x)=1$, we have
\begin{equation*}
\begin{aligned}
\deg\det\mathcal{A}_{l}
&=\sum_{t=0}^{l}(\deg \mathcal{G}_t(x)+(k-1)t).
\end{aligned}
\end{equation*}
Therefore, $\deg\mathcal{A}_{l}-\deg\det\mathcal{A}_{l}=\sum_{t=0}^{l}(\deg \mathsf{w}_t^{(\tau_t)}(x)+(k-1)\tau_t-(k-1)t)$.
Since $\deg \mathsf{w}_t^{(\tau_t)}(x)\leq (n-1)(t-\tau_t)$, we have
\begin{equation*}
\begin{aligned}
\deg\mathcal{A}_{l}-\deg\det\mathcal{A}_{l}\leq \sum_{t=0}^{l}((n-1)(t-\tau_t)+(k-1)\tau_t-(k-1)t).
\end{aligned}
\end{equation*}
Therefore, when $\tau_t=0$, $\max\{\deg\mathcal{A}_l-\deg\det\mathcal{A}_l\}=\frac{1}{2}(n-k)(l^2+l)$.
\hfill$\blacksquare$


\end{document}